\documentclass[twocolumn,tighten,numberedappendix,twocolappendix,trackchanges]{aastex63}

\usepackage{graphicx}
\usepackage{amsmath}
\usepackage{amssymb}
\usepackage{lipsum}
\usepackage{xcolor}

\graphicspath{{./}{plots/}}

\shorttitle{34GHz Deep Field: Stacking Analysis}
\shortauthors{Algera et al.}

\begin{document}

\title{COLDz: Probing Cosmic Star Formation With Radio Free-free Emission}

\author[0000-0002-4205-9567]{Hiddo S. B.\ Algera}
\affiliation{Leiden Observatory, Leiden University, P.O. Box 9513, 2300 RA Leiden, the Netherlands}
\email{algera@strw.leidenuniv.nl}

\author{Jacqueline A.\ Hodge}
\affiliation{Leiden Observatory, Leiden University, P.O. Box 9513, 2300 RA Leiden, the Netherlands}

\author{Dominik A.\ Riechers}
\affiliation{I. Physikalisches Institut, Universit\"{a}t zu K\"{o}ln, Z\"{u}lpicher Straße 77, 50937 K\"{o}ln}

\author{Sarah K.\ Leslie}
\affiliation{Leiden Observatory, Leiden University, P.O. Box 9513, 2300 RA Leiden, the Netherlands}

\author{Ian Smail}
\affiliation{Centre for Extragalactic Astronomy, Durham University, Department of Physics, South Road, Durham, DH1 3LE, UK}

\author{Manuel Aravena}
\affiliation{N\'{u}cleo de Astronom\'{i}a, Facultad de Ingenier\'{i}a y Ciencias, Universidad Diego Portales, Av. Ej\'{e}rcito 441, Santiago, Chile}

\author{Elisabete da Cunha}
\affiliation{International Centre for Radio Astronomy Research, University of Western Australia, 35 Stirling Hwy, Crawley, WA 6009, Australia}
\affiliation{Research School of Astronomy and Astrophysics, Australian National University, Canberra, ACT 2611, Australia}
\affiliation{ARC Centre of Excellence for All Sky Astrophysics in 3 Dimensions (ASTRO 3D)}

\author{Emanuele Daddi}
\affiliation{CEA, IRFU, DAp, AIM, Universit\'e Paris-Saclay, Universit\'e de Paris,  Sorbonne Paris Cit\'e, CNRS, F-91191 Gif-sur-Yvette, France}

\author{Roberto Decarli}
\affiliation{INAF -- Osservatorio di Astrofisica e Scienza dello Spazio di Bologna, via Gobetti 93/3, I-40129, Bologna, Italy}

\author{Mark Dickinson}
\affiliation{NSF's National Optical-Infrared Astronomy Research Laboratory (NOIRLab), 950 North Cherry Avenue, Tucson, AZ 85719, USA}

\author[0000-0003-1436-7658]{Hansung B.\ Gim}
\affiliation{Department of Astronomy, University of Massachusetts Amherst, 710 N Pleasant Street, MA 01003, USA}
\affiliation{Department of Physics, Montana State University, P. O. Box 173840, Bozeman, MT 59717, USA}

\author{Lucia Guaita}
\affiliation{Departamento de Ciencias F\'{i}sicas, Universidad Andr\'{e}s Bello, Fernandez Concha 700, Las Condes, Santiago, Chile}
\affiliation{N\'{u}cleo de Astronom\'{i}a, Facultad de Ingenier\'{i}a y Ciencias, Universidad Diego Portales, Av. Ej\'{e}rcito 441, Santiago, Chile}

\author[0000-0002-6777-6490]{Benjamin Magnelli}
\affiliation{Argelander Institut f\"ur Astronomie, Universit\"at Bonn, Auf dem H\"ugel 71, Bonn, D-53121, Germany}

\author{Eric J.\ Murphy}
\affiliation{National Radio Astronomy Observatory, 520 Edgemont Road, Charlottesville, VA 22903, USA}

\author{Riccardo Pavesi}
\affiliation{Department of Astronomy, Cornell University, Ithaca, New York, 14853, USA}

\author{Mark T.\ Sargent}
\affiliation{Astronomy Centre, Department of Physics and Astronomy, University of Sussex, Brighton, BN1 9QH, UK}
\affiliation{International Space Science Institute (ISSI), Hallerstrasse 6, CH-3012 Bern, Switzerland}

\author{Chelsea E.\ Sharon}
\affiliation{Yale-NUS College, 16 College Avenue West 01-220, 138527 Singapore}

\author{Jeff Wagg}
\affiliation{SKA Observatory, Lower Withington Macclesfield, Cheshire SK11 9DL, UK}

\author{Fabian Walter}
\affiliation{Max-Planck-Institut f\"ur Astronomie, K\"onigstuhl 17, D-69117 Heidelberg, Germany}

\author{Min Yun}
\affiliation{Department of Astronomy, University of Massachusetts, Amherst, MA 01003, USA}

\begin{abstract}
Radio free-free emission is considered to be one of the most reliable tracers of star formation in galaxies. However, as it constitutes the faintest part of the radio spectrum -- being roughly an order of magnitude less luminous than radio synchrotron emission at the GHz frequencies typically targeted in radio surveys -- the usage of free-free emission as a star formation rate tracer has mostly remained limited to the local Universe. Here we perform a multi-frequency radio stacking analysis using deep \emph{Karl G. Jansky} Very Large Array observations at 1.4, 3, 5, 10 and 34\,GHz in the COSMOS and GOODS-North fields to probe free-free emission in typical galaxies at the peak of cosmic star formation. We find that $z\sim0.5 - 3$ star-forming galaxies exhibit radio emission at rest-frame frequencies of $\sim65 - 90\,$GHz that is $\sim1.5 - 2\times$ fainter than would be expected from a simple combination of free-free and synchrotron emission, as in the prototypical starburst galaxy M82. We interpret this as a deficit in high-frequency synchrotron emission, while the level of free-free emission is as expected from M82. We additionally provide the first constraints on the cosmic star formation history using free-free emission at $0.5 \lesssim z \lesssim 3$, which are in good agreement with more established tracers at high redshift. In the future, deep multi-frequency radio surveys will be crucial in order to accurately determine the shape of the radio spectrum of faint star-forming galaxies, and to further establish radio free-free emission as a tracer of high-redshift star formation.

\keywords{galaxies: evolution $--$ galaxies: formation $--$ galaxies: high-redshift $--$ galaxies: star formation}

\end{abstract}

\section{Introduction}
\label{sec:introduction}
One of the major goals in extragalactic astronomy is to constrain the cosmic star formation rate density (SFRD). The SFRD is known to peak between $z\sim1-3$, and then declines rapidly towards the present (e.g., \citealt{madau2014,bouwens2020,leslie2020,katsianis2021,zavala2021}). However, beyond $z\gtrsim3$ star-formation rates have predominantly been measured using rest-frame ultra-violet observations \citep{bouwens2020}. Whilst a powerful tracer of star formation, UV emission is easily attenuated by dust, and may therefore miss an appreciable fraction of the total star formation taking place in the early Universe \citep{casey2018}. In turn, uncertain dust corrections are typically adopted in order to constrain the earliest epochs of cosmic star formation \citep{bouwens2009,bouwens2014,burgarella2013,oesch2013}. Such complications may be circumvented by instead using infrared observations, which probe dust-reprocessed starlight, and as such constrain the fraction of star formation that is dust-obscured (e.g., \citealt{kennicutt1998}). However, with most current infrared facilities it is notoriously difficult to probe beyond the peak of cosmic star formation, due to the limited depth and resolution provided by both ground- and space-based facilities \citep{hodge2020}. In recent years, the highly sensitive Atacama Large Millimeter/submillimeter Array (ALMA) has enabled progress out to higher redshift \citep{bouwens2020,dudzeviciute2020,gruppioni2020,zavala2021}, although its limited field of view makes wide-area far-infrared surveys of star formation highly expensive.

At longer wavelengths, radio emission has provided a powerful tracer of cosmic star formation out to $z\sim5$ (\citealt{novak2017,leslie2020,matthews2021}). This relies on the tight correlation between the radio and far-infrared luminosities of star-forming galaxies, which has been established to hold across a wide range of galaxy types in the local Universe \citep{helou1985,condon1992,yun2001,bell2003}. Low-frequency radio synchrotron emission in star-forming galaxies originates predominantly from the shocks produced by supernovae, and as such forms a delayed tracer of star formation activity ($\sim30-100\,$Myr; \citealt{bressan2002}). However, both at low and high redshift, the far-infrared/radio correlation remains an area of active investigation, with various studies finding that it may be non-linear, change with cosmic time, or depend on galaxy type or physical parameters such as stellar mass \citep{ivison2010b,sargent2010,thomson2014,basu2015,magnelli2015,delhaize2017,read2018,algera2020a,delvecchio2021,molnar2021}. In addition, active galactic nuclei (AGN) may similarly emit at radio wavelengths, and can therefore further bias studies of radio star formation (e.g., \citealt{molnar2018,algera2020a}). Combined with the presently incomplete theoretical underpinning of the far-infrared/radio correlation, the appropriate conversion between radio luminosity and star formation rate in the high-redshift Universe remains not fully understood.

However, the radio regime offers an additional tracer of star formation: at high frequencies ($\nu\gtrsim30\,$GHz), free-free emission is expected to overtake synchrotron radiation as the dominant mechanism generating radio emission \citep{condon1992,murphy2011,tabatabaei2017,querejeta2019}. Free-free emission is produced during the Coulomb interaction of ions and electrons within a dense plasma, and originates directly from the H\,{\normalsize II} regions associated with sites of massive star formation in galaxies. Owing to the short lifetimes of individual H\,{\normalsize II} regions, free-free emission traces star formation on short timescales ($\lesssim10\,$Myr; \citealt{kennicutt2012}), while its long wavelength nature ensures it is mostly insensitive to obscuration by dust. Therefore, free-free emission provides a direct and dust-unbiased tracer of star formation that has been used to calibrate various local tracers \citep{murphy2011,murphy2012}. The clear next step, then, is to investigate this powerful tracer in the early Universe.

Targeting free-free emission at high redshift, however, remains challenging with current radio facilities (e.g., \citealt{thomson2012}). \citet{algera2020c} recently presented a blind survey of free-free emission in high-redshift galaxies, using a 34\,GHz-selected sample identified in deep observations from the \emph{Karl G. Jansky} Very Large Array (VLA) CO Luminosity Density at High Redshift survey (COLD$z$; \citealt{pavesi2018,riechers2019,riechers2020}), in combination with multi-frequency ancillary data. \citet{algera2020c} identified seven star-forming galaxies in these observations with 34\,GHz flux densities dominated by a combination of free-free and synchrotron emission, and as such provided the first blind constraints on free-free emission at high redshift. While limited to a modest sample, they found a good agreement between star formation rates determined from free-free emission and those from canonical tracers such as spectral energy distribution (SED) fitting and the far-infrared/radio correlation.

With present facilities, it remains prohibitively expensive to expand the study of free-free emission at high redshift to significantly larger galaxy samples, and push beyond the bright star-forming population. As an example, a galaxy with a star-formation rate of just $10\,\rm{M}_\odot\,\text{yr}^{-1}$ at $z=1$ ($z=2$) is expected to have a 34\,GHz flux density of $S_{34} \approx 1.0\,\mu\text{Jy}$ ($S_{34} \approx 0.3\,\mu\text{Jy}$), which will remain out of reach for radio telescopes until the advent of the next-generation VLA. However, through a multi-frequency stacking analysis, it is possible to already study the average high-frequency radio emission in normal star-forming galaxies. In this work, we adopt such stacking techniques in combination with the deep radio observations available across the Cosmic Evolution Survey (COSMOS; \citealt{scoville2007}) and the Great Observatories Origins Deep Survey North (GOODS-N; \citealt{giavalisco2004}) in order to provide the first constraints on the nature of free-free emission in representative high-redshift galaxies.

In addition, a multi-frequency stacking analysis simultaneously allows for constraints on the shape of the radio spectra of typical star-forming galaxies. In recent years, a growing amount of evidence has suggested that the radio spectrum of star-forming galaxies may be more complex than the widely adopted combination of just power-law free-free and synchrotron emission. Local Ultra-Luminous Infrared Galaxies (ULIRGs), for example, typically show radio spectra that steepen towards higher frequencies \citep{clemens2008,leroy2011,galvin2018}. At high redshift, the radio spectra of the radio-bright population have similarly been studied in detail, revealing relatively typical synchrotron-dominated spectra at low frequencies (rest-frame $\lesssim 5\,$GHz; \citealt{ibar2010,thomson2014,calistrorivera2017,algera2020a}). However, subsequent follow-up probing higher rest-frame frequencies in starburst galaxies ($\text{SFR}\gtrsim100\,M_\odot\,\text{yr}^{-1}$) indicate their radio spectra might show spectral steepening similar to local ULIRGs \citep{thomson2019,tisanic2019}, which is most readily interpreted as a deficit of free-free emission, or spectral aging of the synchrotron component. Finally, a puzzling component dubbed anomalous microwave emission has been observed in local star-forming regions and galaxies \citep{murphy2015,murphy2020}, occupying a similar frequency range as free-free emission. Deep radio observations, capable of probing rest-frame frequencies $\nu \gtrsim 10\,$GHz are crucial in order to better understand what powers the high-frequency radio emission in galaxies.

The structure of this paper is as follows. In Section \ref{sec:data}, we introduce the radio and ancillary data utilized in this work. In Section \ref{sec:methods}, we detail the stacking analysis and our modelling of the radio spectrum. We present stacked radio spectra of the high-redshift galaxy population in Section \ref{sec:results} and interpret our results in Section \ref{sec:discussion}. Finally, we summarize our findings in Section \ref{sec:conclusion}. Throughout this work, we assume a standard $\Lambda$CDM cosmology, with $H_0=70\,\text{km\,s}^{-1}\text{\,Mpc}^{-1}$, $\Omega_m=0.30$ and $\Omega_\Lambda=0.70$ and adopt a \citet{chabrier2003} initial mass function. The radio spectral index $\alpha$ is further defined as $S_\nu \propto \nu^\alpha$, where $S_\nu$ represents the flux density at frequency $\nu$.

\section{Data}
\label{sec:data}

\subsection{Radio Data}
In this work we combine various sensitive multi-frequency VLA observations across the COSMOS and GOODS-N fields. At the core of our analysis lie the COLD$z$ 34\,GHz continuum observations, which are described in detail in \citet{pavesi2018} and \citet{algera2020c}. These observations combine a deep but small mosaic in the COSMOS field, and a shallower but wider radio map in GOODS-N, following the traditional ``wedding-cake'' design. Briefly, the data in the COSMOS field consist of a 7-pointing mosaic accounting for a total of 93\,hr of on-source time across the VLA D and DnC configurations. The central root-mean-square (RMS) noise in the map is $1.3\,\mu\text{Jy\,beam}^{-1}$, and the mosaic covers a field of view of $9.6\,\text{arcmin}^2$. The COSMOS data were designed to overlap with a prominent $z=5.3$ protocluster, the brightest member of which is individually detected in the 34\,GHz observations (AzTEC-3; \citealt{algera2020c}). We investigate the radio properties of the additional cluster members in Appendix \ref{app:protocluster}, while we focus on the unbiased sample of lower redshift galaxies in the foreground in the remainder of this work. The GOODS-N field was observed for 122\,hr on-source across the VLA D, D$\rightarrow$DnC, DnC and DnC$\rightarrow$C configurations. The resulting 57-pointing mosaic spans an area of $51\,\text{arcmin}^2$, with a typical RMS of $5.3\,\mu\text{Jy\,beam}^{-1}$. In addition, a single deep pointing within the mosaic, designed to overlap with the NOEMA 3\,mm line observations in \citet{decarli2014}, probes down to $3.2\,\mu\text{Jy\,beam}^{-1}$. Both mosaics reach a typical resolution of $2''$ - $2\farcs5$, which is large enough that most continuum detections remain unresolved \citep{algera2020c}, allowing for the cleanest measurement of their flux densities.

Deep ancillary radio data are crucial in order to accurately constrain the shape of the radio spectrum in star-forming galaxies. A full description of the available radio data across the COSMOS and GOODS-N fields is given in \citet{algera2020c}, which we summarize in Table 1, as well as briefly below. In the COSMOS field, we employ sensitive observations at 3 and 10\,GHz from the COSMOS-XS survey \citep{algera2020b,vandervlugt2020}, which fully cover the COLD$z$ footprint. These data reach a typical RMS sensitivity of $0.53\,\mu\text{Jy\,beam}^{-1}$ and $0.41\,\mu\text{Jy\,beam}^{-1}$ at 3 and 10\,GHz, respectively, and as such are a factor of $\sim10\times$ more sensitive toward radio synchrotron emission from star-forming galaxies than the COLD$z$ 34\,GHz observations, after a spectral scaling with a typical $\alpha=-0.70$ \citep{condon1992}. At both frequencies, the COSMOS-XS observations attain a typical resolution of $\sim2\farcs0$, similar to that of the 34\,GHz data.

\begin{deluxetable}{lllcc}

\label{tab:radio}
\tablecaption{Properties of the radio data utilized in COSMOS (upper four rows) and GOODS-N (lower).}
\tablehead{
    \colhead{$\nu_\mathrm{obs}$} &
	\colhead{$\mathrm{RMS}(\nu_\mathrm{obs})$} & 
	\colhead{$\mathrm{RMS}(1.4\,\mathrm{GHz})$} & 
	\colhead{$\theta_M \times \theta_m$} & 
	\colhead{Ref}
}

\startdata
GHz & $\mu\mathrm{Jy\,beam}^{-1}$ & $\mu\mathrm{Jy\,beam}^{-1}$ & $\mathrm{arcsec}^2$ \\
\tableline
1.4 & 1.8 & 1.8 & $1.35\times1.21$ & 1 \\
3 & 0.53 & 0.90 & $2.21\times1.86$ & 2 \\
10 & 0.41 & 1.6 & $2.26\times1.98$ & 2 \\
34 & 1.3 & 12 & $2.70\times2.41$ & 3 \\
\tableline
1.4 & 2.2 & 2.2 & $1.60\times1.60$ & 4 \\
5 & 3.5 & 8.5 & $1.47\times1.42$ & 5 \\
10 & 1.5 & 5.9 & $2.00\times2.00$ & 6 \\
34 & 5.3 & 49 & $2.19\times1.84$ & 3
\enddata

\tablerefs{[1] Algera et al. (in prep.); [2] \citet{vandervlugt2020}; [3] \citet{algera2020c}; [4] \citet{owen2018}; [5] \citet{gim2019}; [6] \citet{murphy2017}.} 
\tablecomments{(1) Frequency; (2) RMS noise at native frequency; (3) RMS noise scaled to 1.4\,GHz with $\alpha = -0.70$; (4) Beam size; (5) References}
\end{deluxetable}

To constrain the low-frequency radio emission of the sources individually detected in the 34\,GHz data across COSMOS, \citet{algera2020c} adopted the 1.4\,GHz observations from \citet{schinnerer2007,schinnerer2010} which reach a typical RMS of $12\,\mu\text{Jy\,beam}^{-1}$. However, in this work we utilize new, more sensitive VLA observations at 1.4\,GHz covering the COSMOS-XS and COLD$z$ footprints as part of the COSMOS-XL survey (PI: Algera). These observations will be fully described in a forthcoming publication (Algera et al., in preparation), but we briefly summarize their key properties here. The COSMOS field was observed in a single 1.4\,GHz pointing for a total of 26.5\,hr in the VLA A-configuration, centered on R.A. $10^\text{h}00^\text{m}20.7^\text{s}$, Decl. $+02^\circ32'52\farcs6$. These observations were taken between 20 Oct 2019 and 24 Feb 2021 as part of VLA programs 19A-370 and 20A-370. The data were calibrated using the standard VLA pipeline, and imaging was performed in CASA 5.7.1 via {\sc{tclean}}. We adopted a multi-frequency synthesis algorithm with \emph{nterms} = 2 to account for the large fractional bandwidth, and used w-projection to account for the non-coplanarity of baselines. The data were Briggs-weighted with a robust parameter of 0.5. Prior to the primary beam correction, the median RMS within $20\%$ of the primary beam sensitivity equals $1.8\,\mu\text{Jy\,beam}^{-1}$. As a result, these 1.4\,GHz observations are roughly $7$ times deeper than the existing VLA observations at 1.4\,GHz across the COLD$z$ footprint. In addition, they are also roughly $7$ times deeper than the 34\,GHz data across COSMOS, assuming a standard spectral index of $\alpha = -0.70$ \citep{condon1992}. \\


The GOODS-N field similarly benefits from a wealth of ancillary radio observations. We make use of the 1.4\,GHz map from \citet{owen2018}, which reaches a typical RMS-noise of $2.2\,\mu\text{Jy\,beam}^{-1}$ in the pointing center, at a resolution of $1\farcs6$. In addition, \citet{gim2019} covered the GOODS-N field with two VLA pointings at 5\,GHz. Their data reach an RMS-noise of $3.5\,\mu\text{Jy\,beam}^{-1}$, and attain a resolution of $1\farcs5$. Furthermore, \citet{murphy2017} imaged the GOODS-N field at 10\,GHz in a single VLA pointing, covering approximately 75\% of the COLD$z$ footprint. At their native resolution of $0\farcs22$, the 10\,GHz observations reach an RMS sensitivity of $0.57\,\mu\text{Jy\,beam}^{-1}$. However, in this work we make use of the tapered maps provided by \citet{murphy2017} to ensure that we accurately capture all the flux of the (stacked) radio sources. The 10\,GHz map tapered to $1''$ ($2''$) reaches a central RMS of $1.1\,\mu\text{Jy\,beam}^{-1}$ ($1.5\,\mu\text{Jy\,beam}^{-1}$). For our analysis, we adopt the 10\,GHz map with a $2''$ taper to better match the resolution of the ancillary radio maps.

The archival radio data in GOODS-N are of a higher relative sensitivity than the 34\,GHz map, when scaled with a fixed spectral index of $\alpha=-0.70$. At 1.4\,GHz, the \citet{owen2018} radio map is roughly $20\times$ deeper, while at 5 and 10\,GHz, the radio images from \citet{gim2019} and \citet{murphy2017} are, respectively $6\times$ and $8\times$ more sensitive. As such, we expect to be limited by the S/N at 34\,GHz in our analysis. Nevertheless, the COLD$z$ continuum data provide crucial high-frequency constraints on the radio spectra of star-forming galaxies, and form the foundation of this work.

At the typical resolution of our radio data of $1\farcs5 - 2\farcs0$, we do not expect to resolve (stacks of) star-forming galaxies, which are typically sub-arcsecond in size in the $\mu$Jy regime \citep{murphy2017,bondi2018,cotton2018,jimenez-andrade2019,jimenez-andrade2021,muxlow2020}. In addition, galaxies are expected to become increasingly compact towards higher radio frequencies, which form the focus of this work \citep{murphy2017,thomson2019}. In turn, we do not expect to resolve out any emission when measuring radio flux densities, allowing for unbiased spectral index measurements. However, the resolution of our radio data is additionally high enough that any effects of source blending are negligible.

\subsection{Optical/FIR Data}
We employ deep optical and infrared observations across the COSMOS and GOODS-N fields, to serve as prior positional information for our stacking analysis. In the COSMOS field, we make use of the $z^{++}YJHK_s$-selected COSMOS2015 catalog from \citet{laigle2016}, which compiles data spanning UV to far-infrared wavelengths. \citet{laigle2016} additionally use the SED-fitting code {\sc{LePhare}} \citep{ilbert2009} to determine photometric redshifts, stellar masses and star-formation rates for all entries in the catalog. In total, $1158$ galaxies from COSMOS2015 fall within $20\%$ of the COLD$z$/COSMOS primary beam sensitivity. 

In order to derive useful and unbiased constraints on the radio properties of star-forming galaxies via a stacking analysis, it is necessary to assess the completeness of the input sample. The mass completeness of the COSMOS2015 catalog is determined by \citet{laigle2016}, who estimate the catalog to be $90\%$ complete above stellar masses of $10^9, 10^{9.5}$ and $10^{10}\,M_\odot$ out to $z\lesssim 1.3, z\lesssim 2.3$ and $z\lesssim 4.0$, respectively.\footnote{These completeness limits were determined for the UltraVISTA ``ultra-deep'' stripes, with which the COLD$z$ 34\,GHz observations overlap in their entirety.}

Across the GOODS-N field, we employ the photometry compiled in the 3D-HST catalog \citep{brammer2012,skelton2014}. Source detection for 3D-HST was performed in a combined F125+F140W+F160W image, with additional photometry being performed in 22 filters spanning the $U-$band to \emph{Spitzer}/IRAC CH4. These observations are further extended by \citet{momcheva2016}, who determine the redshift for all 3D-HST entries by combining broadband photometry with \emph{HST}/GRISM spectroscopic observations. In addition, \citet{momcheva2016} determine dust-corrected star-formation rates by including information from \emph{Spitzer}/MIPS $24\,\mu$m observations. The mass completeness of the 3D-HST catalog has been assessed by \citet{tal2014}. They determine the catalog to be roughly 90\% complete above stellar masses of $10^{9}\,M_\odot$, $10^{9.5}\,M_\odot$ and $10^{10.0}\,M_\odot$ out to $z \lesssim 1.8$, $z \lesssim 2.5$ and $z\lesssim 3.0$, respectively. In total, $14{,}313$ galaxies included in the 3D-HST catalog fall within the footprint of the COLD$z$/GOODS-N observations, within 20\% of the peak primary beam sensitivity.

\section{Methods}
\label{sec:methods}

\subsection{Radio Stacking}
\label{sec:methods_stacking}

In this work we employ a stacking analysis in order to investigate the shape of the radio spectrum of typical star-forming galaxies between observed-frame $1.4 - 34\,$GHz. To this end, we create small cutouts of $51\times51$ pixels ($25\farcs5\times25\farcs5$ at 34\,GHz) around galaxy positions identified in optical/near-infrared (NIR) imaging within the various radio maps, and co-add them together to gain a census of their average radio emission. As star-forming sources are expected to be faint at high radio frequencies, a large number of sources are required to be averaged together in order to obtain a clear detection even in the stacks. This, in turn, requires co-adding sources across a relatively wide range in redshift. In this work, we therefore stack in luminosity as opposed to flux density, in order to fairly combine sources across different cosmic epochs. For a source at redshift $z$, with a flux density $S_\nu$ at observed-frame frequency $\nu$, we probe a luminosity of

\begin{align}
    L_{\nu'} = \frac{4\pi D_L(z)^2}{1+z} S_\nu \ ,
\end{align}

\noindent where $\nu' = \nu(1+z)$. However, to ensure we probe the same rest-frame frequency for all sources in a given redshift bin, we scale the flux density to probe $\overline{\nu}' = \nu(1+\overline{z})$, where $\overline{z}$ is the median redshift in the bin, prior to stacking. Since this rest-frame frequency is probed at an observed-frame frequency of $\nu(1+\overline{z})/(1+z)$ for a source at redshift $z$, and $S_\nu \propto \nu^\alpha$, this implies that

\begin{align}
    L_{\overline{\nu}'} = \frac{4\pi D_L(z)^2}{1+\overline{z}}\left( \frac{1+\overline{z}}{1+z}\right)^{1+\alpha} S_\nu \ .
    \label{eq:lumstacking}
\end{align}

As such, an assumption on the spectral index must be made a priori. However, given that there are by definition an equal number of sources in the redshift bin with $z>\overline{z}$ as there are with $z<\overline{z}$, any uncertainty induced as a result of this spectral scaling tends to be small. In addition, with the typical $\alpha=-0.70$ we assume \citep{condon1992}, the exponent $1+\alpha$ in Equation \ref{eq:lumstacking} constitutes only a relatively shallow power, further minimizing any uncertainty induced by the luminosity-stacking. We have verified that adopting any reasonable value of $\alpha$ between $-0.30$ and $-1.10$ does not change the corresponding stacked luminosity density within more than a few per cent. \\

As we adopt relatively wide redshift bins in Section \ref{sec:results_optical}, with a typical $\Delta z / \overline{z} \approx 1$, we briefly discuss how this may affect our analysis. Invariably, any (redshift) evolution in the galaxy population within individual bins is averaged over in our stacking procedure. Most of the galaxies analyzed in this work reside on the so-called star formation main sequence \citep{brinchmann2004,noeske2007}, and, given that the normalization of the main sequence increases with redshift \citep{speagle2014,schreiber2015}, the typical star formation rate within a single bin is similarly expected to increase towards high redshift.\footnote{As an example, the typical SFR of a main-sequence galaxy with $M_\star = 10^{10.3}\,M_\odot$ (the median mass in our highest redshift bin; Section \ref{sec:results_optical}) evolves by a factor of 3 between $z = 1$ and $z = 3$ \citep{schreiber2015}.} Given the strong correlation between star formation rate and radio luminosity, galaxies towards the upper redshift range of the bins are typically more luminous. Our stacking analysis, in turn, provides the typical radio luminosity across a galaxy sample spanning a relatively wide range in star formation rate. While any variations in, for example, the far-infrared/radio correlation as a function of SFR (a non-linear correlation as advocated by e.g., \citealt{molnar2021}) are averaged across in our analysis, the resulting stacks accurately describe the typical radio luminosities of the underlying sample.

If, however, the shape of the radio spectrum depends on star formation rate, this would introduce a varying radio $k$-correction across individual bins. While with present data this cannot be investigated at high radio frequencies, a recent study by \citet{an2021} finds no evidence for an SFR-dependent spectral index between $1.3 - 3\,$GHz for a large galaxy sample in COSMOS. As such, it is unlikely that our wide bins conceal significant redshift evolution, and hence our stacked radio luminosities should not be significantly affected. \\

In this work we adopt both a median and a mean-stacking analysis, for different purposes. We review the advantages and disadvantages of either method in Appendix \ref{app:stacking}, and briefly summarize our choice here. In Section \ref{sec:results}, we set out to determine the typical radio spectrum of individually undetected star-forming galaxies. In this case, we adopt a median stacking analysis, in order to obtain a radio spectrum that is representative of the underlying galaxy population, and less susceptible to contamination from AGN.

One caveat that applies when adopting the median, however, is that its interpretation is complicated in the presence of noise. In particular, as shown by \citet{white2007}, the stacked median tends to be ``boosted'' with respect to the true sample median when the noise in the radio maps used for stacking is similar to, or exceeds, the typical flux density of the underlying galaxy population (which a priori is unknown). We investigate the effect of median boosting by testing our stacking routine on realistically generated mock sources in Appendix \ref{app:stacking}, and calculate the deboosting factors required to accurately compare stacked flux densities. These correction factors are typically largest at low S/N, and may reach up to $f_\text{boost} \approx 2 - 2.5$ for (simulated) low luminosity galaxies at 34\,GHz. Throughout this work, all quoted stacked flux densities and luminosities are corrected for the effects of median boosting, unless explicitly stated otherwise.\\

In Section \ref{sec:discussion}, we seek to place constraints on the cosmic star formation history through free-free emission. In this case, we adopt a mean stacking analysis, as we are not interested in the median star formation rate of individual galaxies, but instead in the \emph{total} amount of star-formation occurring in a given redshift slice. This total simply constitutes the product of the number of sources being stacked and their mean-stacked star formation rate. A further advantage of the mean is that, compared to the median, it is straightforward to interpret, and is robust in the presence of noise, i.e., it does not require any deboosting corrections. However, in contrast, AGN contamination is more likely to affect the mean stacks. As such, we remove sources identified as likely radio AGN from the stacks in Section \ref{sec:results}.

In our mean stacking approach, we additionally need to ensure that the background level in the stacks is not significantly affected by bright neighboring galaxies. As such, we treat individually detected radio sources separately from the undetected population and  stack in the residual radio images, from which all bright radio sources have been removed (following, e.g., \citealt{magnelli2015}). These residual maps were created using {\sc{PyBDSF}} \citep{mohanrafferty2015}, using the appropriate detection threshold to match that of the parent catalogs.\footnote{As an example, source detection in the 34\,GHz radio maps was performed with a $3\sigma$ peak threshold \citep{algera2020c}, while the 3 and 10\,GHz images in COSMOS used a $5\sigma$ threshold \citep{vandervlugt2020}.} The radio-detected sources are added in a posteriori, via

\begin{align}
    \langle L_{\nu'} \rangle = \frac{N_\text{undet} \times \langle L_{\nu',\text{undet}} \rangle + \sum_{i=1}^{N_\text{det}} L_{\nu',\text{det},i}}{N_\text{undet} + N_\text{det}} \ .
    \label{eq:meanstacking}
\end{align}

\noindent Here $N_\text{det}$ ($N_\text{undet}$) is the number of detected (undetected) radio sources used in the stacking, $L_{\nu',\text{undet}}$ the stacked luminosity of the undetected sources at an average frequency $\nu'$, and $L_{\nu',\text{det},i}$ the luminosity of the $i^\text{th}$ individually detected galaxy, at the same frequency $\nu'$. We ensure the same rest-frame frequency $\nu'$ is probed for the stack and the detections -- namely the aforementioned $\nu' = \nu(1+\overline{z})$, where $\overline{z}$ represents the median redshift across the detected and undetected sources combined -- by adopting $\alpha=-0.70$ and scaling $L_{\nu'}$ accordingly. The error on the combined luminosity $\langle L_{\nu'} \rangle$ incorporates both the error on the stacked luminosity, and the uncertainty on the individually detected sources. 

We perform photometry on the stacks using \textsc{PyBDSF}, following \citet{algera2020a}, which fits a 2D Gaussian to any significant emission in the center of the stack. We pass an estimate of both the background mean and RMS in the stacked cutout by simultaneously stacking random locations within the radio images, which are mostly devoid of sources. Unless stated otherwise, we employ a $3\sigma$ detection threshold for the stacks. If no detection is found at this significance, a $3\sigma$ upper limit is adopted instead. When applying the median boosting corrections (Appendix \ref{app:stacking}), we also propagate the spread of the recovered mock source fluxes into the error on the true deboosted flux densities. This effectively takes into account variation among the input sample into the final uncertainty, with the key benefit that we do not explicitly need to perform a bootstrap analysis on the real stacks: as bootstrapping by construction involves duplicating sources within the input distribution, the error in the bootstrapped stack increases with respect to the original sample.\footnote{For a large number of cutouts $N$ with identical noise properties, the RMS in the bootstrapped stack is larger than that in the original stack by a factor of $\sqrt{2}$.} As our high-frequency stacks typically have modest significance ($\text{S/N} \approx 3 - 5$), bootstrapping results in both an increased number of non-detections and a larger median boosting correction, and as such constitutes a suboptimal approach in the low-S/N regime. We emphasize that this procedure requires the distribution of mock sources to match the true galaxy distribution with a high accuracy, which we ensure to be the case in Appendix \ref{app:stacking}.

\subsection{Radio Spectral Decomposition}
\label{sec:decomposition}

The radio spectrum of star-forming galaxies is frequently modelled as the sum of two power-law processes: synchrotron and free-free emission (e.g., \citealt{condon1992}). The former has a power-law slope with a typical value of $\alpha_\text{NT} = -0.85$ \citep{niklas1997,murphy2012}, though observations of high-redshift sources indicate substantial scatter ($\sigma_{\alpha_\mathrm{NT}}=0.3-0.5$; \citealt{smolcic2017a,calistrorivera2017}). Free-free emission, on the other hand, has a well-known and nearly flat spectral index of $\alpha_\text{FF}=-0.10$ \citep{condon1992,murphy2011}.\footnote{Note the different spectral indices adopted: synchrotron emission has a typical slope of $\alpha_\mathrm{NT} = -0.85$, while, at low frequencies, the overall radio spectrum has a typical slope of $\alpha \approx -0.70$ owing to the additional contribution of free-free emission \citep{condon1992,smolcic2017a}.} The radio spectrum can therefore be written as

\begin{align}
    L_{\nu'} = L_{\nu_0'} \left[ \left(1-f_{\nu_0'}^\text{th} \right) \left( \frac{\nu'}{\nu_0'}\right)^{\alpha_\text{NT}} + f_{\nu_0'}^\text{th} \left( \frac{\nu'}{\nu_0'}\right)^{-0.1} \right] \ ,
    \label{eq:spectrum}
\end{align}

\noindent where the thermal fraction $f_{\nu_0'}^\text{th}$ represents the relative contribution of free-free emission to the total radio emission at rest-frame frequency $\nu_0'$ (e.g., \citealt{tabatabaei2017}). In turn, the radio spectrum can be fully characterized by three parameters: $f_{\nu_0'}^\text{th}$, $\alpha_\text{NT}$ and an overall normalization $L_{\nu_0'}$. We adopt $\nu_{0}' = 1.4\,$GHz in this work, which is the rest-frame frequency where the thermal fraction is typically defined \citep{condon1992,tabatabaei2017}. We run a Monte Carlo Markov Chain (MCMC) fitting routine to determine these parameters, as well as accurate uncertainties. We adopt flat priors on $f_\text{th}$ and $L_{\nu_0'}$, and allow unphysical negative values in order to assess whether a thermal component is preferred by the fitting. We further adopt a Gaussian prior on $\alpha_\text{NT}$ centered on a mean value of $-0.85$, with a spread of $\sigma=0.30$. This spread is smaller than the $\sigma=0.50$ adopted by \citet{algera2020c}, as they model the radio spectra of individual sources, whereas in this work we consider only stacked radio spectra. For a stacked sample of sources, the average synchrotron slope is expected to regress towards the typical value of $\alpha_\text{NT}\approx-0.85$, justifying the assumption of a narrower prior. For further details on the spectral fitting routine, we refer the reader to \citet{algera2020c}. 

Once the thermal fraction is known, star formation rates can be determined from the observed free-free luminosity. We adopt the calibration from \citet{murphy2012}, adapted for a Chabrier IMF, which is given by

\begin{align}
\begin{split}
    \left( \frac{\text{SFR}_\text{FF}}{M_\odot\,\text{yr}^{-1}} \right) = 4.3 &\times 10^{-28} \left( \frac{T_e}{10^4\,\text{K}} \right)^{-0.45} \left( \frac{\nu}{\text{GHz}}\right)^{0.10} \\ 
    &\times \left( \frac{f_\text{th}(\nu') L_{\nu'}}{\text{erg\,s}^{-1}\,\text{Hz}^{-1}}\right) \ .
    \label{eq:ffsfr}
\end{split}
\end{align}

\noindent Here $T_e = 10^4\,$K is the electron temperature of the H{\,\normalsize II} regions, upon which the star formation rate weakly depends. We refer the reader to \citet{murphy2012} and \citet{querejeta2019} for a detailed discussion of the assumptions going into the calibration of free-free emission as a star formation rate tracer.

Throughout this work, we compare our results with a simple but widely used model for the radio spectrum, namely that of prototypical starburst galaxy M82. M82 has a star-formation rate of $\text{SFR}\sim10-20\,M_\odot\,\text{yr}^{-1}$ \citep{forsterschreiber2003}, which is similar to the typical SFRs of the galaxy population analyzed in this work (Section \ref{sec:discussion_sfrs}). The radio spectrum of M82 can be well described by a combination of free-free and synchrotron emission \citep{condon1992}, and as such forms the natural comparison to our high-redshift galaxy sample. We will in the following refer to an ``M82-like'' model for the radio spectrum as having a thermal fraction of $f_\text{th}(1.4\,\text{GHz}) = 0.1$ \citep{condon1992}, and a synchrotron spectral index of $\alpha_\text{NT} = -0.85$ \citep{niklas1997,murphy2012}. While more complex forms of the radio spectrum -- in particular in very luminous starburst galaxies -- have been observed (e.g., \citealt{galvin2018}), the M82-like model remains the most commonly assumed shape of the radio spectrum in the absence of wide multi-frequency radio coverage (e.g., \citealt{murphy2017,tabatabaei2017,klein2018,penney2020}). In addition, the M82-like radio spectrum provides a good description of the radio spectra of the 34\,GHz-selected star-forming galaxies analyzed by \citet{algera2020c}, across an identical frequency range as explored in this work ($1.4 - 34\,$GHz).

\section{Results}
\label{sec:results}

\subsection{Free-free Emission in Optically-selected Galaxies}
\label{sec:results_optical}

\begin{figure}[t]
    \centering
    \includegraphics[width=0.5\textwidth]{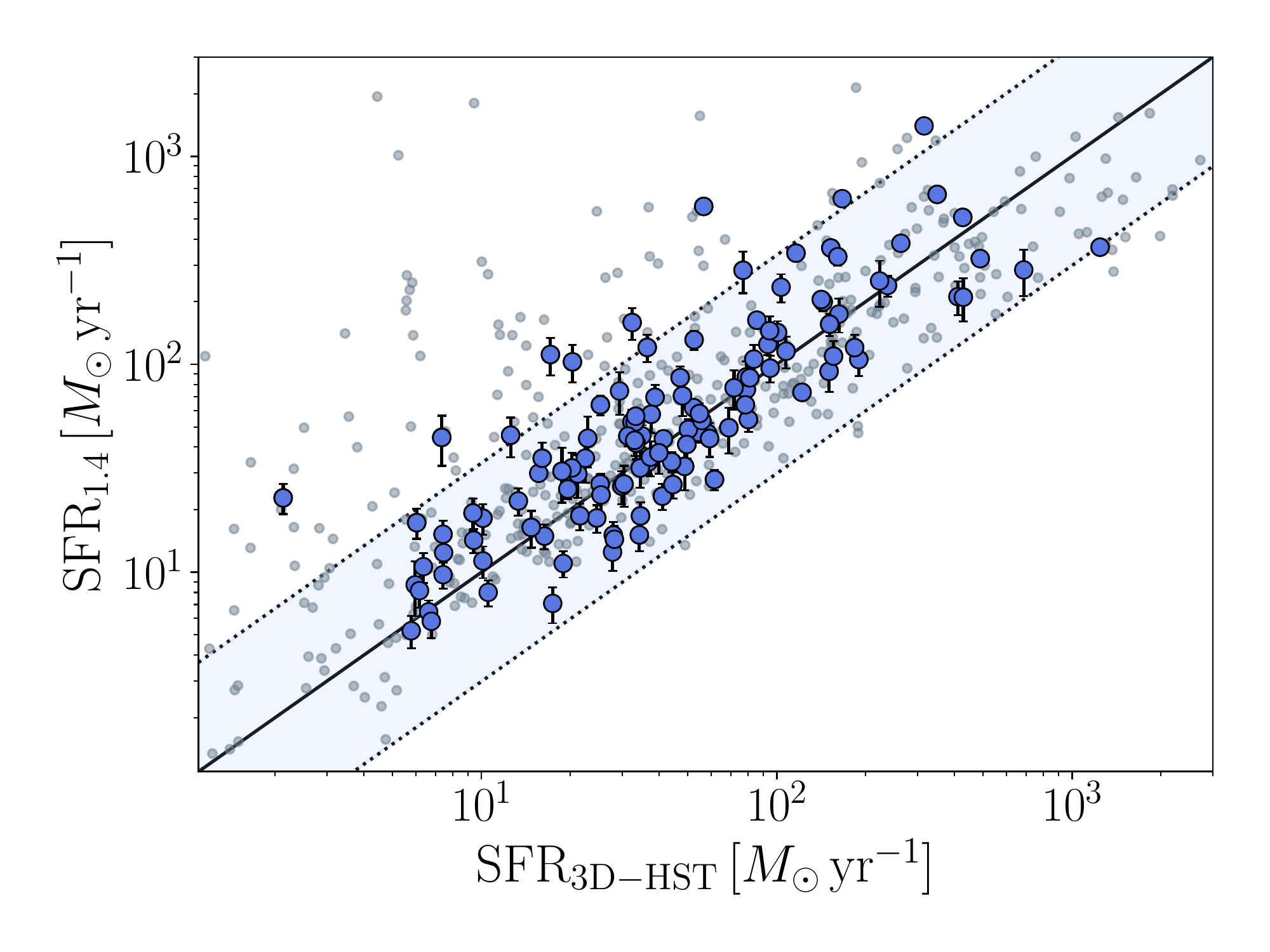}
    \caption{Star formation rates at 1.4\,GHz, determined via the far-infrared/radio correlation from \citet{delvecchio2021}, as a function of the optical/near-infrared star formation rates from the 3D-HST catalog. The full sample of matches between the \citet{owen2018} and 3D-HST catalog is shown via the grey circles, while the galaxies falling within any of the bins defined in Table \ref{tab:binning} are shown as larger blue points. The blue shaded region indicates the $2.5\sigma$ scatter about the far-infrared/radio correlation (0.53\,dex). Sources with large radio SFRs placing them above this region are identified as radio AGN and are excluded from the stacking analysis.}
    \label{fig:goodsn_agn}
\end{figure}

\begin{figure*}[t]
    \centering
    \includegraphics[width=0.495\textwidth]{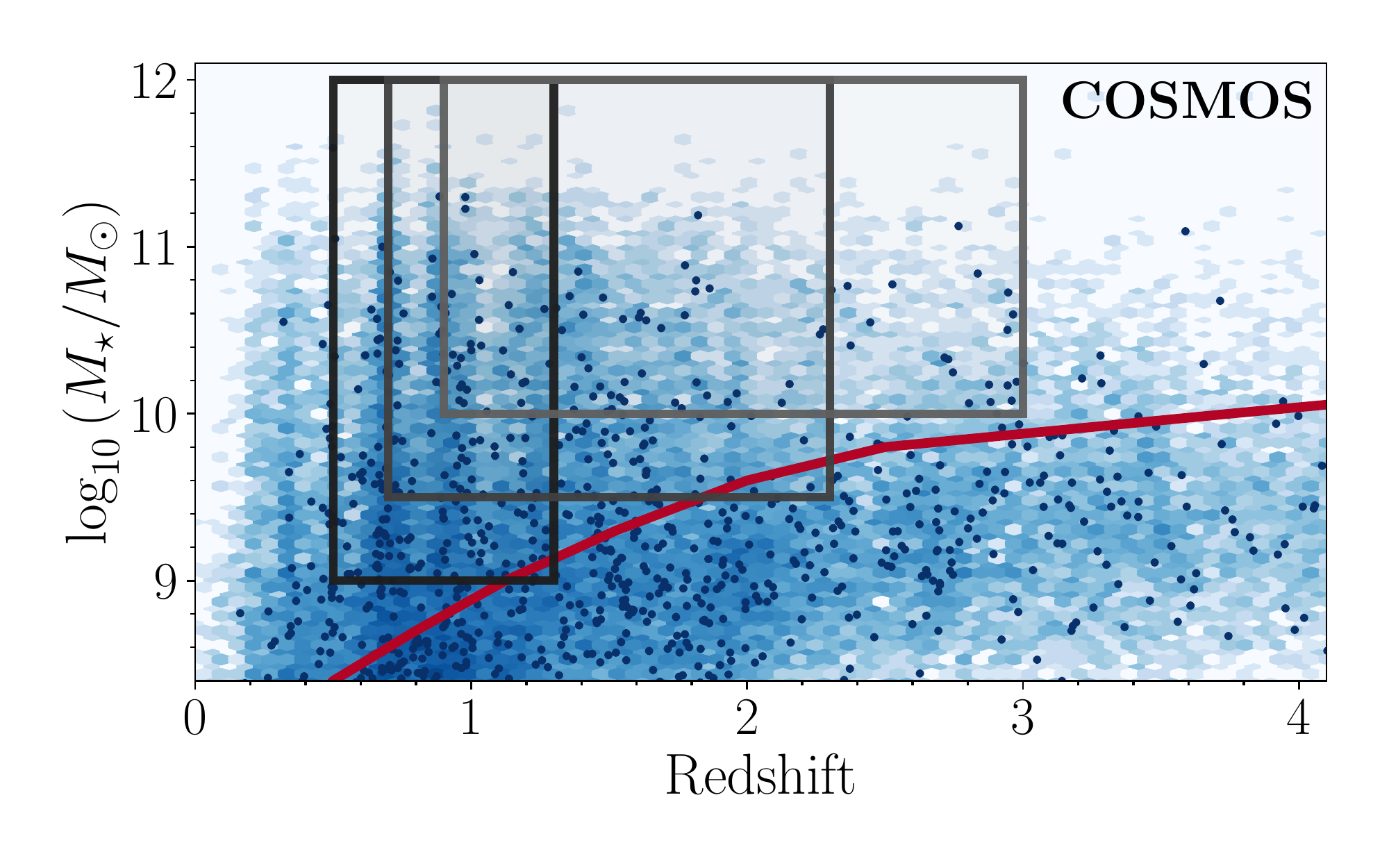}
    \includegraphics[width=0.495\textwidth]{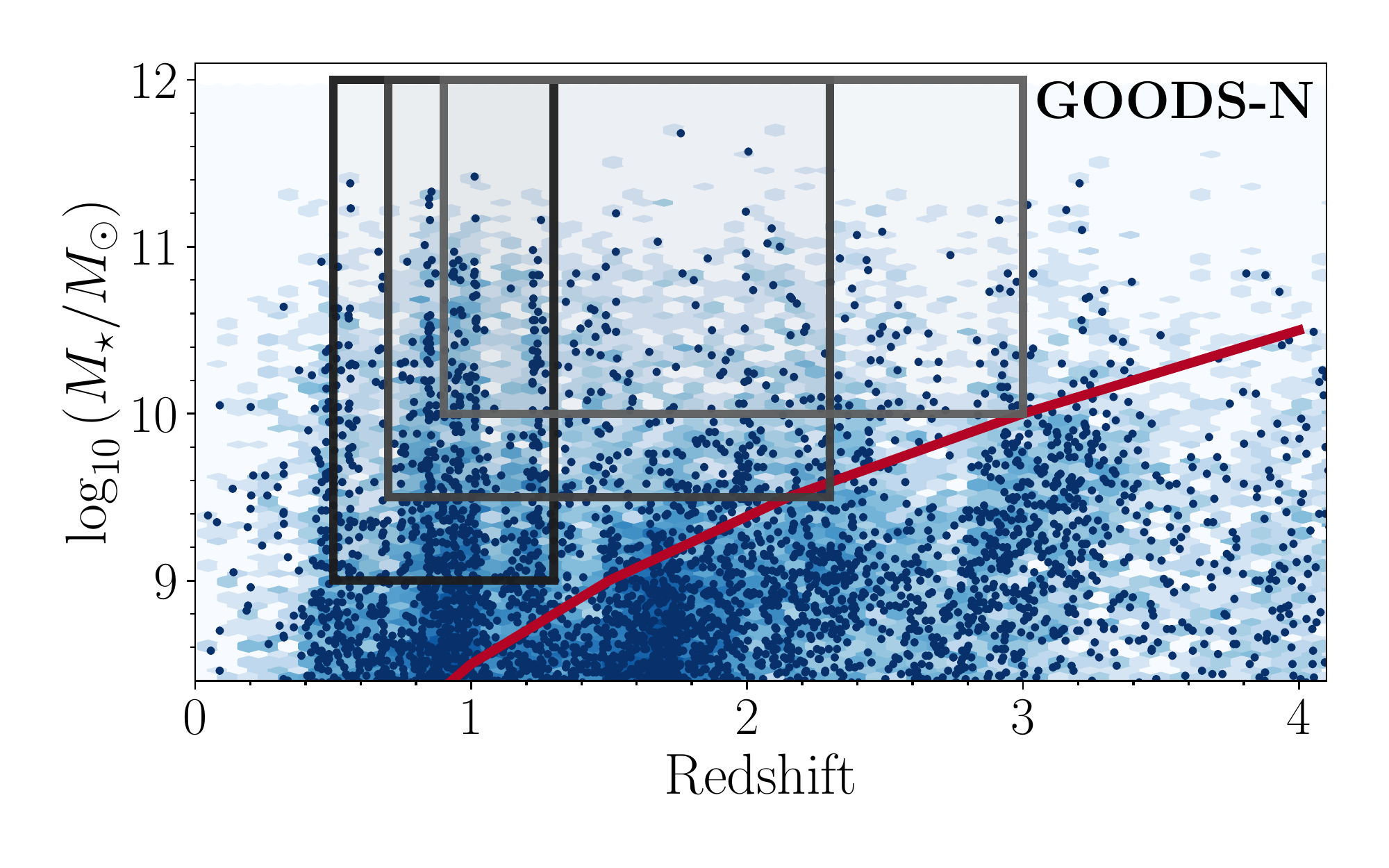}
    \caption{Adopted binning in the COSMOS (left) and GOODS-N fields (right). The background histogram illustrates the distribution of sources in the stellar mass vs.\ redshift plane across the full COSMOS2015 and 3D-HST catalogs, while the individual points mark galaxies within the field of view of the 34\,GHz observations, or the combined field of view of the 10 and 34\,GHz observations in GOODS-N. The solid lines indicate the 90\% mass completeness as determined by \citet{laigle2016} and \citet{tal2014} for COSMOS and GOODS-N, respectively. The grey rectangles indicate the adopted mass-complete binning, which is identical between the two fields.}
    \label{fig:binning}
\end{figure*}

\begin{figure*}[t]
    \centering
    \includegraphics[width=0.495\textwidth]{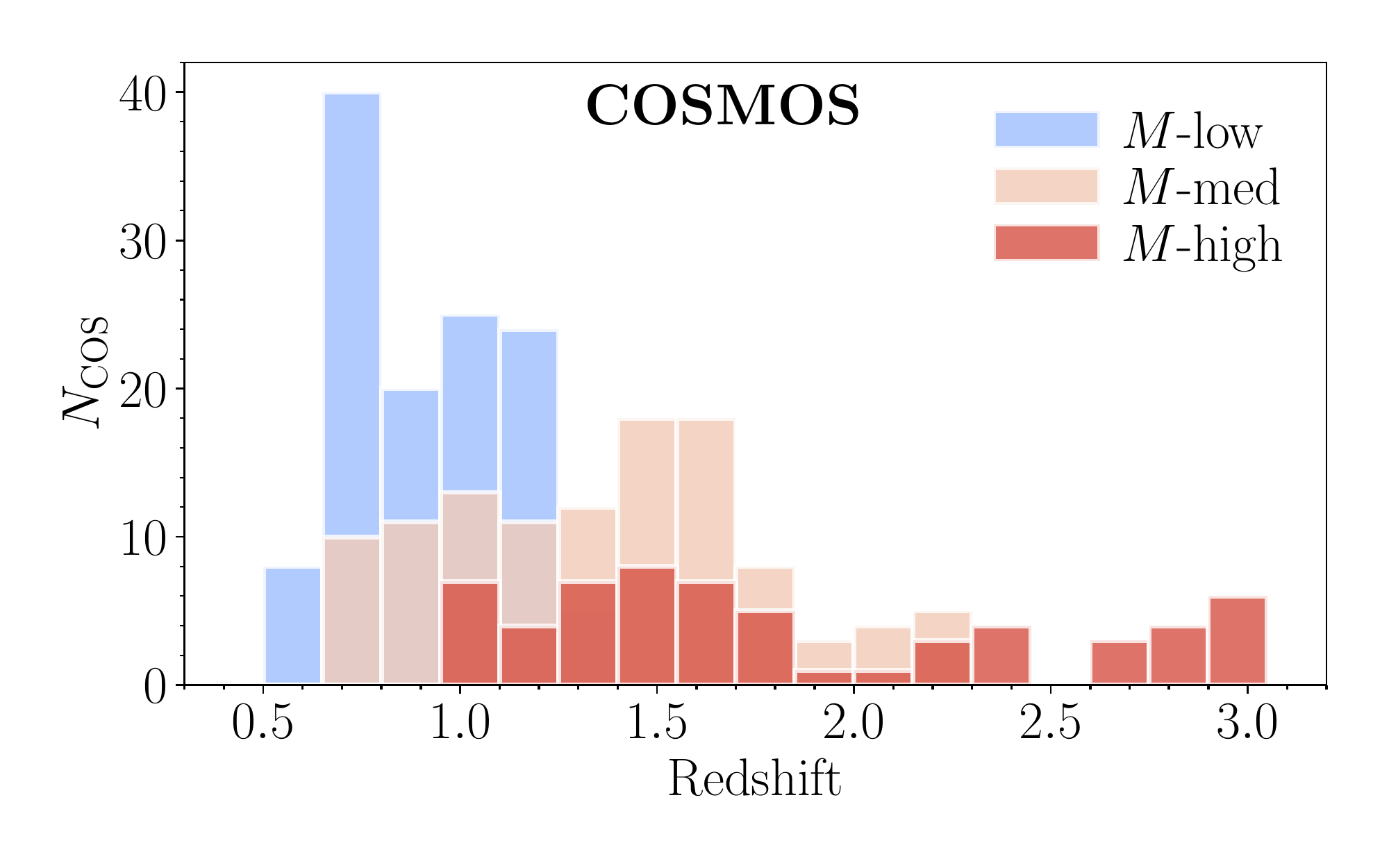}
    \includegraphics[width=0.495\textwidth]{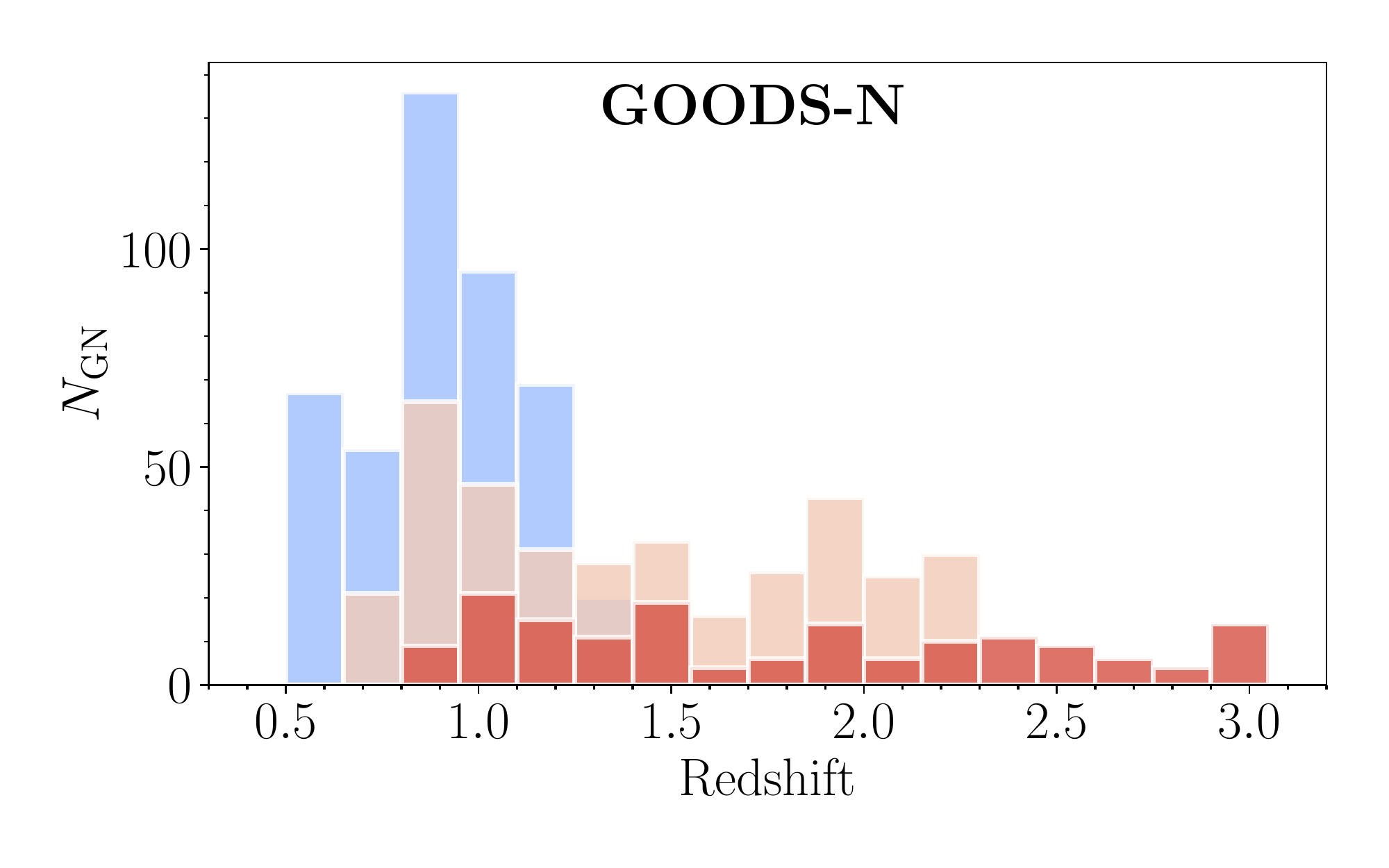}
    \caption{Redshift distributions of the star-forming galaxies in the COSMOS (left) and GOODS-N fields (right). Sources are subdivided into the three partially overlapping mass-complete bins defined in Table \ref{tab:binning}.}
    \label{fig:redshift_dist}
\end{figure*}

\begin{deluxetable}{cccccccc}

\label{tab:binning}
\tablecaption{Mass-complete bins constructed from the COSMOS2015 and 3D-HST catalogs.}
\tablehead{
    \colhead{Bin} &
	\colhead{$z_1$} & 
	\colhead{$z_2$} & 
	\colhead{$\overline{z}$} &
	\colhead{$>$\,$\log M_\star$} & 
	\colhead{$\langle\log {M}_\star\rangle$} &
	\colhead{$N_\text{COS}$} & 
	\colhead{$N_\text{GN}$}
}

\startdata
$M$-low & 0.5 & 1.3 & 0.92 & 9.0 & 9.5 & 122 & 441 (642) \\
$M$-med & 0.7 & 2.3 & 1.36 & 9.5 & 9.9 & 113 & 364 (553) \\
$M$-high & 0.9 & 3.0 & 1.60 & 10.0 & 10.3 & 60 & 159 (242) \\
\enddata

\tablecomments{(1) Bin identifier; (2), (3), (4) Lower, upper and median redshift of the bin, combining both fields; (5), (6) Minimum and median stellar mass; (7), (8) Number of sources in the bin in COSMOS and GOODS-N (combining the 10 \& 34\,GHz areas; $N_\text{GN}$ within the full COLD$z$ field of view is included in parentheses).}
\end{deluxetable}

We now set out to constrain the average radio spectrum of representative star-forming galaxies, which are typically not individually detected even in deep radio observations. In COSMOS, we adopt prior galaxy positions from the COSMOS2015 catalog, and remove all galaxies that can be cross-matched to a radio AGN identified in the COSMOS-XS survey within $0\farcs9$ \citep{algera2020b}. These AGN were identified through their offset from the far-infrared/radio correlation derived by \citet{delhaize2017} at a significance of $>2.5\sigma$, with far-infrared luminosities having been derived with SED fitting code {\sc{magphys}} \citep{dacunha2008,dacunha2015}. While radio-quiet AGN are as such not explicitly removed from this sample, these tend to show radio emission similar to the star-forming population \citep{delvecchio2017,algera2020a,algera2020b}. Additionally, the fraction of radio-quiet AGN decreases strongly towards faint radio flux densities \citep{smolcic2017b,algera2020b}, thereby making it unlikely that such AGN significantly bias our analysis. We additionally ensure the remaining galaxies are star-forming based on their position in the $\text{NUV}-r$, $r-J$ color-color diagram, following \citet{ilbert2013}. 

In the GOODS-N field, we adopt prior positions from the 3D-HST catalog. We limit ourselves to the area where the 10\,GHz observations from \citet{murphy2017} and the COLDz 34\,GHz continuum data overlap, and perform the stacking analysis at four frequencies (1.4, 5, 10 and 34\,GHz). For the radio-detected population in GOODS-N, however, there is no a piori available information on whether the radio emission is likely originating from star formation or from an AGN. In order to still exclude radio AGN, we therefore first cross-match the \citet{owen2018} 1.4\,GHz catalog with the 3D-HST catalog, adopting a matching radius of $0\farcs9$. We subsequently determine star formation rates from the radio luminosities at 1.4\,GHz -- adopting $\alpha = -0.70$ for the required $K$-corrections -- via (following \citealt{delhaize2017}):

\begin{align}
    \left(\frac{\mathrm{SFR}_\mathrm{1.4\,GHz}}{M_
    \odot\,\mathrm{yr}^{-1}}\right) = 10^{-24}\times10^{q_\mathrm{IR}} \left(\frac{L_{1.4}}{\mathrm{W\,Hz}^{-1}}\right) \ .
    \label{eq:synchrotron_sfr}
\end{align}

\noindent Here $q_\mathrm{IR}$ is the parameterization of the far-infrared/radio correlation (e.g., \citealt{helou1985,bell2003}). We adopt the recent mass-dependent far-infrared/radio correlation $q_\mathrm{IR}(z,M_\star)$ from \citet{delvecchio2021} (their Equation 5), who additionally determine a typical scatter about the correlation of $\sigma_{q_\mathrm{IR}} = 0.21\,$dex. We compare the radio star formation rates with the optical/NIR (OIR) star formation rates from the 3D-HST catalog in Figure \ref{fig:goodsn_agn}. Galaxies are identified as radio AGN when their radio star formation rates exceed the OIR values, after accounting for a scatter of $2.5\times\sigma_{q_\mathrm{IR}}$ about the far-infrared/radio correlation. We additionally ensure the remaining galaxies are star-forming based on their position in the UVJ-diagram (e.g., \citealt{williams2009}), adopting the rest-frame magnitudes provided by \citet{skelton2014}.

We subsequently divide the galaxies into three wide, mass-complete redshift bins, removing all sources for which the cutout does not fully lie within the COLD$z$ footprint. The low ($>10^{9}\,\rm{M}_\odot$), medium ($>10^{9.5}\,\rm{M}_\odot$) and high-mass ($>10^{10}\,\rm{M}_\odot$) bins (henceforth referred to as $M$-low, $M$-med and $M$-high, respectively) extend to the maximum redshift where both the COSMOS2015 and 3D-HST catalogs are complete. This, in turn, allows for a direct comparison of the results across both fields in Section \ref{sec:discussion}. We show the bins for both fields on the stellar mass versus redshift plane in Figure \ref{fig:binning}, and present the redshift distributions of the sources within the bins in Figure \ref{fig:redshift_dist}. In addition, the precise binning we adopt and the number of sources per bin are given in Table \ref{tab:binning}. We note that the bins partially overlap in redshift, and are therefore not fully independent of one another. However, in order to obtain a stacked high-frequency detection at sufficient S/N, adopting wide bins is essential. As a result, across both fields $33-41\%$ of sources are in common between adjacent bins $M$-low and $M$-med or bins $M$-med and $M$-high.

\begin{figure*}[!t]
    \centering
    \vspace*{-1cm}
    \includegraphics[width=1.0\textwidth]{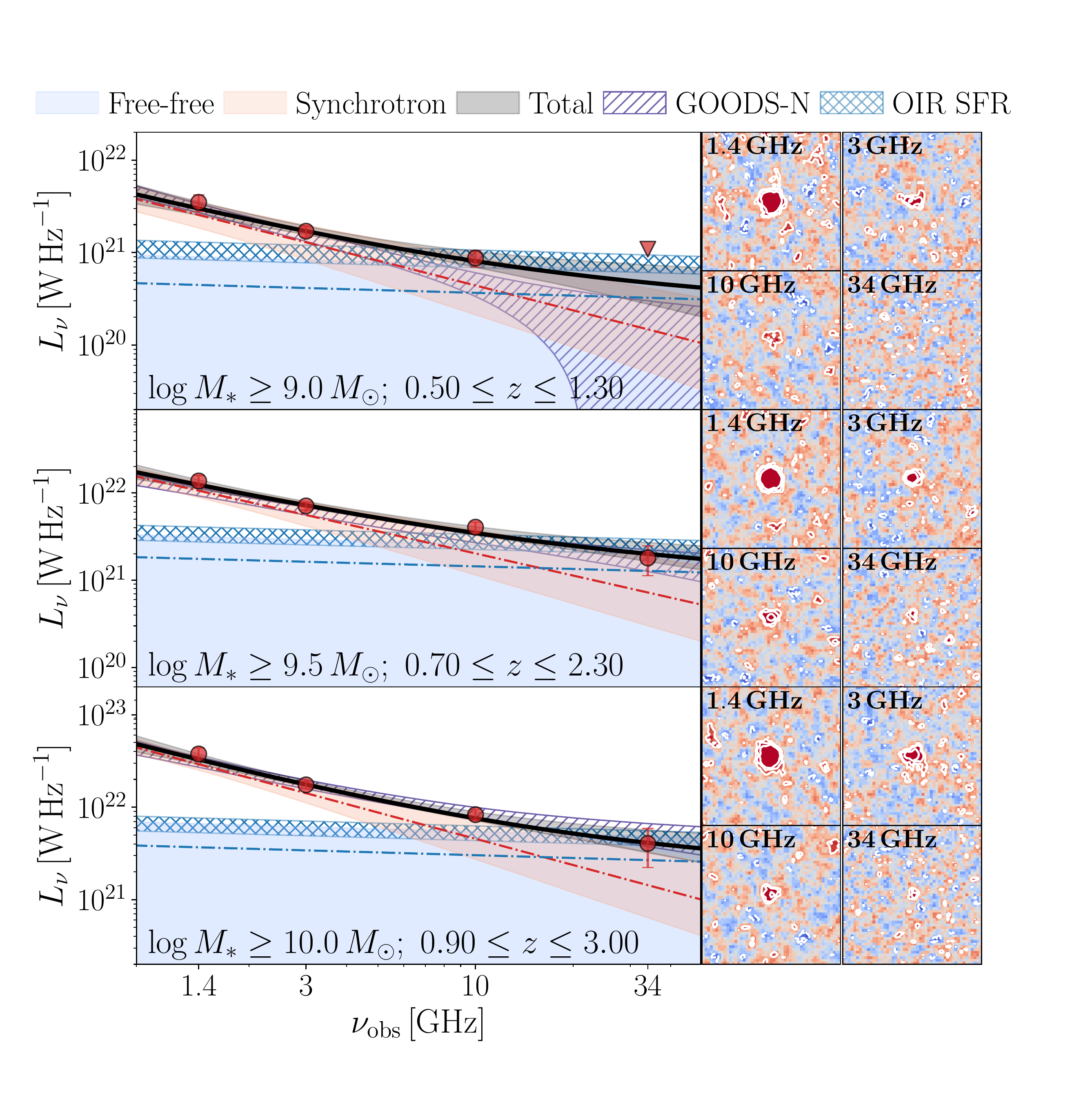}
    \vspace*{-1.2cm}
    \caption{Median-stacked radio spectra (left) and stacked cutouts ($51\times51$ pixels; right) in the COSMOS field, for the three different mass-complete bins highlighted in Table \ref{tab:binning}. The blue and red shaded regions in the radio spectra show the $1\sigma$ confidence intervals on the fitted free-free and synchrotron emission, respectively. The black line and grey shading comprises the total fitted emission, and the hatched blue region represents the predicted free-free luminosity given the typical SFR derived from the optical/infrared data for the galaxies. The purple hatched region represents the best fit to the stacked radio spectrum in GOODS-N (Figure \ref{fig:stacked_goodsn}), and is shown to allow a direct visual comparison of the fields. In the stacks, contours are shown at the $\pm2,\pm3$ and $\pm5\sigma$ levels, where $\sigma$ is the RMS in the stack. Negative contours are indicated via dashed lines, and the color scale runs from $-3\sigma$ to $+3\sigma$. High-frequency radio emission at 34\,GHz is detected in two out of three bins, at a significance of $4.0\sigma$ and $3.0\sigma$.}
    \label{fig:stacked_cosmos}
\end{figure*}

\begin{figure*}[!t]
    \centering
    \vspace*{-1.0cm}
    \includegraphics[width=1.0\textwidth]{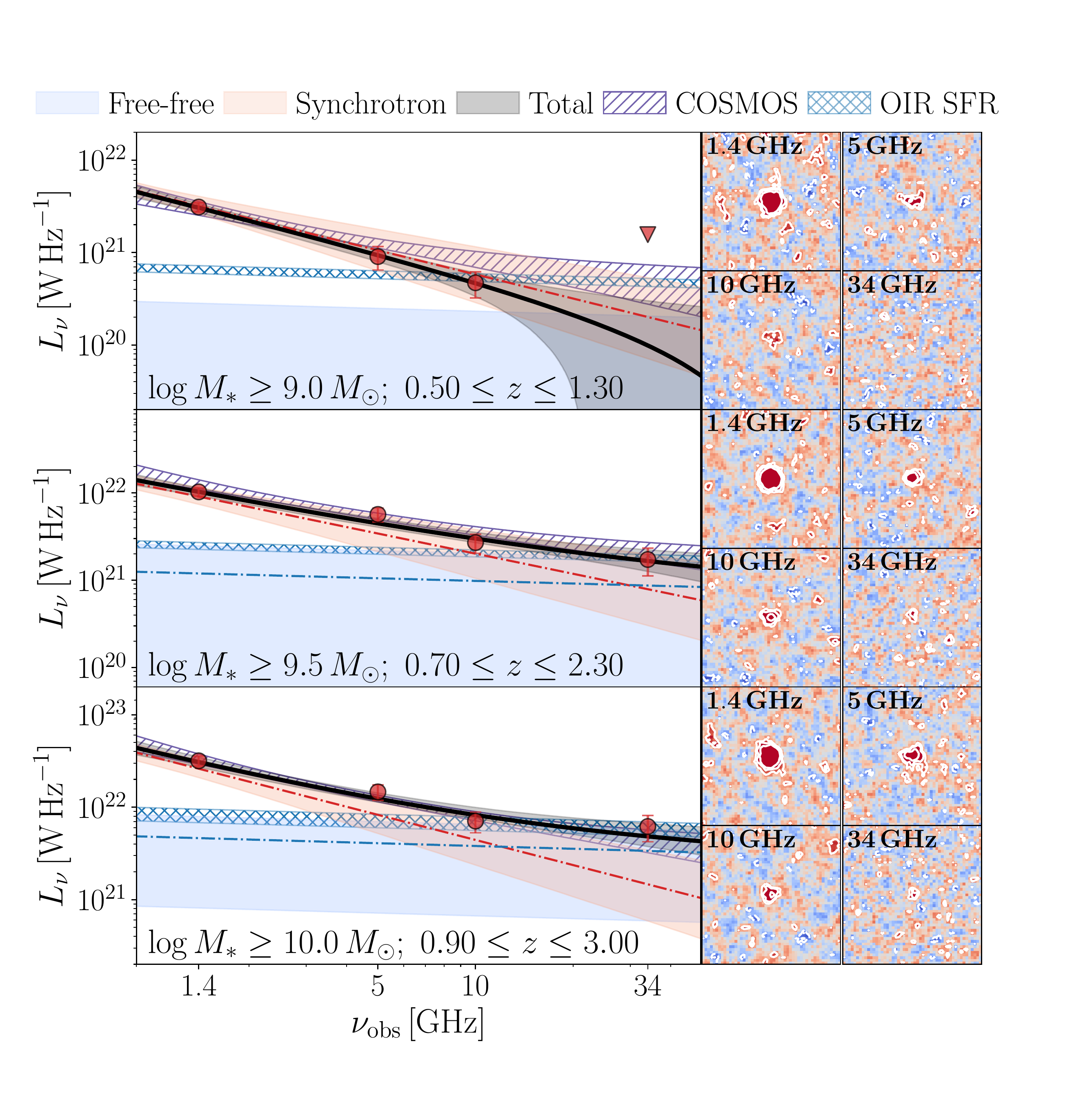}
    \vspace*{-1.2cm}
    \caption{Similar to Figure \ref{fig:stacked_cosmos}, now showing the median-stacked radio spectra (left) and stacked cutouts (right) in mass-complete bins across the GOODS-N field. We detect stacked 34\,GHz continuum emission in two out of three bins, at a significance of $3.4\sigma$ and $4.0\sigma$.}
    \label{fig:stacked_goodsn}
\end{figure*}

We show the median stacked cutouts and the corresponding multi-frequency radio spectra for the COSMOS and GOODS-N fields in Figures \ref{fig:stacked_cosmos} and \ref{fig:stacked_goodsn}, respectively. In COSMOS, we detect stacked 34\,GHz continuum emission at $4.0\sigma$ and $3.0\sigma$ significance in bins $M$-med and $M$-high, respectively, while for bin $M$-low we can only place a $3\sigma$ upper limit. In the GOODS-N field, we detect stacked 34\,GHz emission at $3.4\sigma$ and $4.0\sigma$ significance in bins $M$-med and $M$-high, respectively, while similarly no significant emission is detected in $M$-low. The stacked luminosities in COSMOS and GOODS-N, as well as the adopted deboosting factors, are presented in Table \ref{tab:stacking}. The values for both fields are in good agreement, verifying that similar galaxy populations are probed in COSMOS and GOODS-N.

We additionally show the free-free luminosity expected from star formation in Figures \ref{fig:stacked_cosmos} and \ref{fig:stacked_goodsn}, and compare this to the stacked luminosity density at 34\,GHz. We adopt the star formation rates derived from spectral energy distribution fitting of OIR data from the COSMOS2015 and 3D-HST catalogs by \citet{laigle2016} and \citet{momcheva2016}, respectively, which account for a potential contribution from dust-obscured star formation via deep \emph{Spitzer}/MIPS $24\,\mu$m observations. We subsequently convert the OIR SFRs to the expected free-free luminosity $\langle L_{34}^\mathrm{OIR}\rangle$ by inverting Equation \ref{eq:ffsfr}. We then calculate the OIR-predicted thermal fraction as $f_\text{th}^\mathrm{OIR}\left(34\,\text{GHz}\right) = \langle L_{34}^\text{OIR} \rangle / L_{34}^\text{obs}$, that is, as the ratio of the expected free-free luminosity and the observed $34\,$GHz luminosity. Following this procedure, we predict OIR thermal fractions of $f_\text{th}^\text{OIR} = 0.9 - 1.4$ across the four bins with stacked 34\,GHz detections in COSMOS and GOODS-N, with a mean value of $f_\text{th}^\text{OIR} = 1.1_{-0.2}^{+0.4}$.\footnote{While in practice the thermal fraction cannot exceed unity, we quote the formal errors on the OIR predicted value which combines the uncertainty on $\langle L_{34} \rangle$ and the spread on the OIR SFRs.} The two low-mass bins, at which no emission is detected at 34\,GHz at the $3\sigma$ level, only provide lower limits of $f_\text{th}^\text{OIR} > 0.3 - 0.7$. As such, based on a comparison with the optical-infrared star formation rates, we expect the observed luminosity at 34\,GHz to be dominated by free-free emission. We discuss this further in Section \ref{sec:discussion_sfrs}, where we analyze the two fields jointly.


\movetabledown=5cm

\begin{rotatetable*}
\begin{deluxetable*}{ccccccccc|cc}

\label{tab:stacking}
\tablecaption{Stacked luminosities and deboosting factors in COSMOS (upper three rows) and GOODS-N (lower).}
\tablehead{
    \colhead{Bin} &
	\colhead{$f_\text{boost}(1.4)$} & 
	\colhead{$L_{1.4}$} & 
	\colhead{$f_\text{boost}(3\,\rm{or}\,5)$\tablenotemark{a}} &
	\colhead{$L_{3\,\rm{or}\,5}$\tablenotemark{a}} & 
	\colhead{$f_\text{boost}(10)$} &
	\colhead{$L_{10}$} & 
	\colhead{$f_\text{boost}(34)$} &
	\colhead{$L_{34}$} & 
	\colhead{$f_\text{th}(34)$\tablenotemark{b}} & 
	\colhead{$\alpha_\text{NT}$\tablenotemark{b}}
}

\startdata
$-$ & $-$ & $10^{21}\,\rm{W\,Hz}^{-1}$ & $-$ & $10^{21}\,\rm{W\,Hz}^{-1}$ & $-$ & $10^{21}\,\rm{W\,Hz}^{-1}$ & $-$ & $10^{21}\,\rm{W\,Hz}^{-1}$ & $-$ & $-$ \\  
\hline
$M$-low & $0.88\pm0.13$ & $3.5\pm0.7$ & $0.96\pm0.10$ & $1.7\pm0.2$ & $1.03\pm0.16$ & $0.9\pm0.2$ & $-$ & $<1.1$ & $<0.86$ & $-0.89_{-0.29}^{+0.25}$ \\
$M$-med & $0.79\pm0.10$ & $13.7\pm2.1$ & $0.90\pm0.09$ & $7.1\pm0.7$ & $0.99\pm0.13$ & $4.0\pm0.6$ & $1.64\pm0.46$ & $1.8\pm0.7$ & $0.51_{-0.51}^{+0.24}$ & $-0.84_{-0.26}^{+0.24}$ \\
$M$-high & $0.73\pm0.10$ & $37.6\pm6.0$ & $0.87\pm0.07$ & $17.4\pm1.6$ & $0.90\pm0.10$ & $8.3\pm1.1$ & $1.32\pm0.41$ & $4.1\pm1.8$ & $0.48_{-0.46}^{+0.24}$ & $-0.94_{-0.25}^{+0.24}$ \\
\hline
$M$-low & $1.21\pm0.10$ & $3.1\pm0.3$ & $1.64\pm0.25$ & $0.9\pm0.3$ & $1.90\pm0.33$ & $0.5\pm0.1$ & $-$ & $<1.6$ & $<0.73$ & $-0.87_{-0.28}^{+0.84}$ \\
$M$-med & $1.01\pm0.10$ & $10.3\pm1.1$ & $1.26\pm0.17$ & $5.7\pm1.0$ & $1.35\pm0.20$ & $2.7\pm0.7$ & $2.43\pm0.45$ & $1.7\pm0.6$ & $0.41_{-0.68}^{+0.32}$ & $-0.77_{-0.26}^{+0.32}$ \\
$M$-high & $0.88\pm0.10$ & $31.8\pm4.0$ & $0.92\pm0.11$ & $14.6\pm2.8$ & $1.07\pm0.15$ & $7.1\pm1.8$ & $1.67\pm0.32$ & $6.2\pm2.0$ & $0.54_{-0.45}^{+0.23}$ & $-0.90_{-0.25}^{+0.23}$
\enddata

\tablenotetext{a}{Deboosting factors and luminosities are given at 3\,GHz for COSMOS and 5\,GHz for GOODS-N; see text.}
\tablenotetext{b}{The errors on $f_\text{th}$ and $\alpha_\text{NT}$ are degenerate (Figure \ref{fig:stacked_params}; see also \citealt{algera2020c}). We here quote the one-dimensional uncertainties, but adopt the full posterior probability distributions in this work to accurately propagate the uncertainties.}

\tablecomments{(1) Bin identifier, matching that in Table \ref{tab:binning}; (2), (4), (6), (8) Boosting factors at 1.4, 3 or 5, 10 and 34\,GHz; (3), (5), (7), (9) Median-stacked deboosted luminosities at observed-frame 1.4, 3 or 5, 10 and 34\,GHz; (10) Thermal fraction at observed-frame 34\,GHz; (11) Synchrotron spectral index.}

\end{deluxetable*}
\end{rotatetable*}


\section{Discussion}
\label{sec:discussion}

\subsection{Radio Star-formation Rates}
\label{sec:discussion_sfrs}

\begin{figure*}[t]
    \centering
    \includegraphics[width=0.8\textwidth]{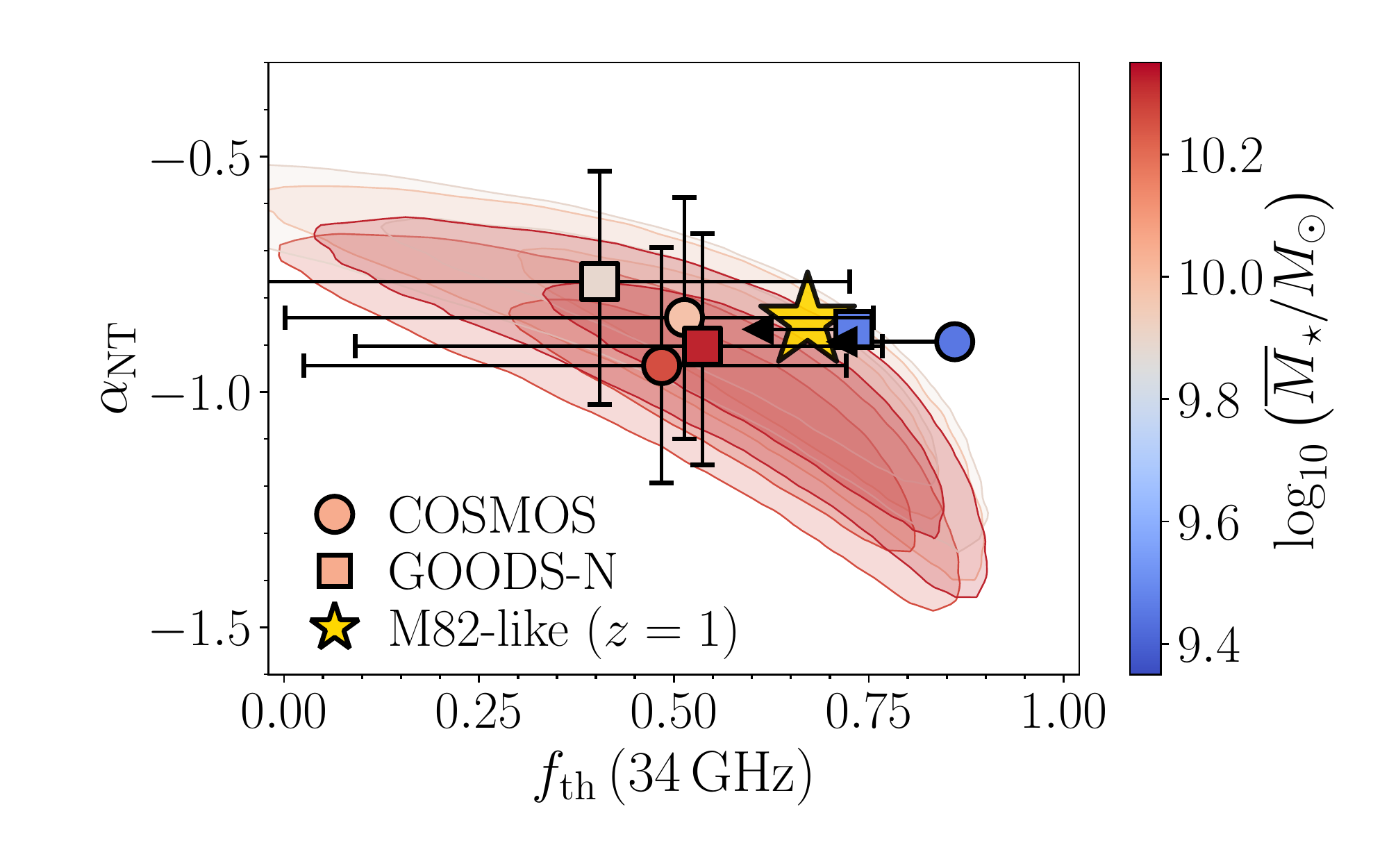}
    \vspace*{-0.5cm}
    \caption{Synchrotron spectral index versus thermal fraction at observed-frame 34\,GHz for the COSMOS (circles) and GOODS-N (squares) median stacks. Points are colored by their median stellar mass, and the 34\,GHz thermal fraction and synchrotron slope for an M82-like radio spectrum at $z=1$ are indicated via the yellow star. The shaded contours represent the $1\sigma$ (dark) and $2\sigma$ (light) confidence intervals on the spectral parameters, which are somewhat degenerate at this low S/N. In case of a non-detection at 34\,GHz, we place an upper limit on the thermal fraction at the median $\alpha_\text{NT}$ comprising 84\% of the sampled values for $f_\text{th}$ (equivalent to a $1\sigma$ upper limit). The recovered thermal fractions are a factor of $\sim1.5 - 2$ lower than predicted from an M82-like radio spectrum.}
    \label{fig:stacked_params}
\end{figure*}

In the previous Section, we sampled the radio spectrum of star-forming galaxies in COSMOS and GOODS-N through a multi-frequency stacking technique. We now jointly analyze the results across these fields, and investigate the nature of synchrotron and free-free emission in faint star-forming galaxies.

At rest-frame frequencies $\nu\gtrsim30\,$GHz, radio free-free emission is expected to dominate the radio spectrum (e.g., \citealt{condon1992,klein2018}). To test this, we decompose the stacked multi-frequency radio spectra in COSMOS and GOODS-N into their synchrotron and free-free components, using the fitting routine outlined in Section \ref{sec:methods}. The decomposed radio spectra are shown in Figures \ref{fig:stacked_cosmos} and \ref{fig:stacked_goodsn}, and the resulting fitted synchrotron spectral indices $\alpha_\text{NT}$ and thermal fractions $f_\text{th}$ are shown in Figure \ref{fig:stacked_params} and are additionally tabulated in Table \ref{tab:stacking}. At typical rest-frame frequencies of $\nu' \approx 65 - 90\,$GHz, we recover relatively low thermal fractions of $f_\text{th} \approx 0.4 - 0.5$. For comparison, a simple extrapolation of the M82 radio SED predicts a thermal fraction of $f_\text{th} \approx 0.7$ in this frequency regime, while the OIR predicted thermal fractions imply even larger values of $f_\text{th} \approx 1$. Accordingly, the fitted thermal fractions appear to be a factor of $\sim1.5-2$ lower than expected. In addition, in four out of the six bins, the thermal fraction at 34\,GHz is consistent with zero within $1\sigma$. In these bins, the combined fit to the radio spectrum is in turn primarily composed of the power-law synchrotron component.

\begin{figure*}[t]
    \centering
    \includegraphics[width=0.49\textwidth]{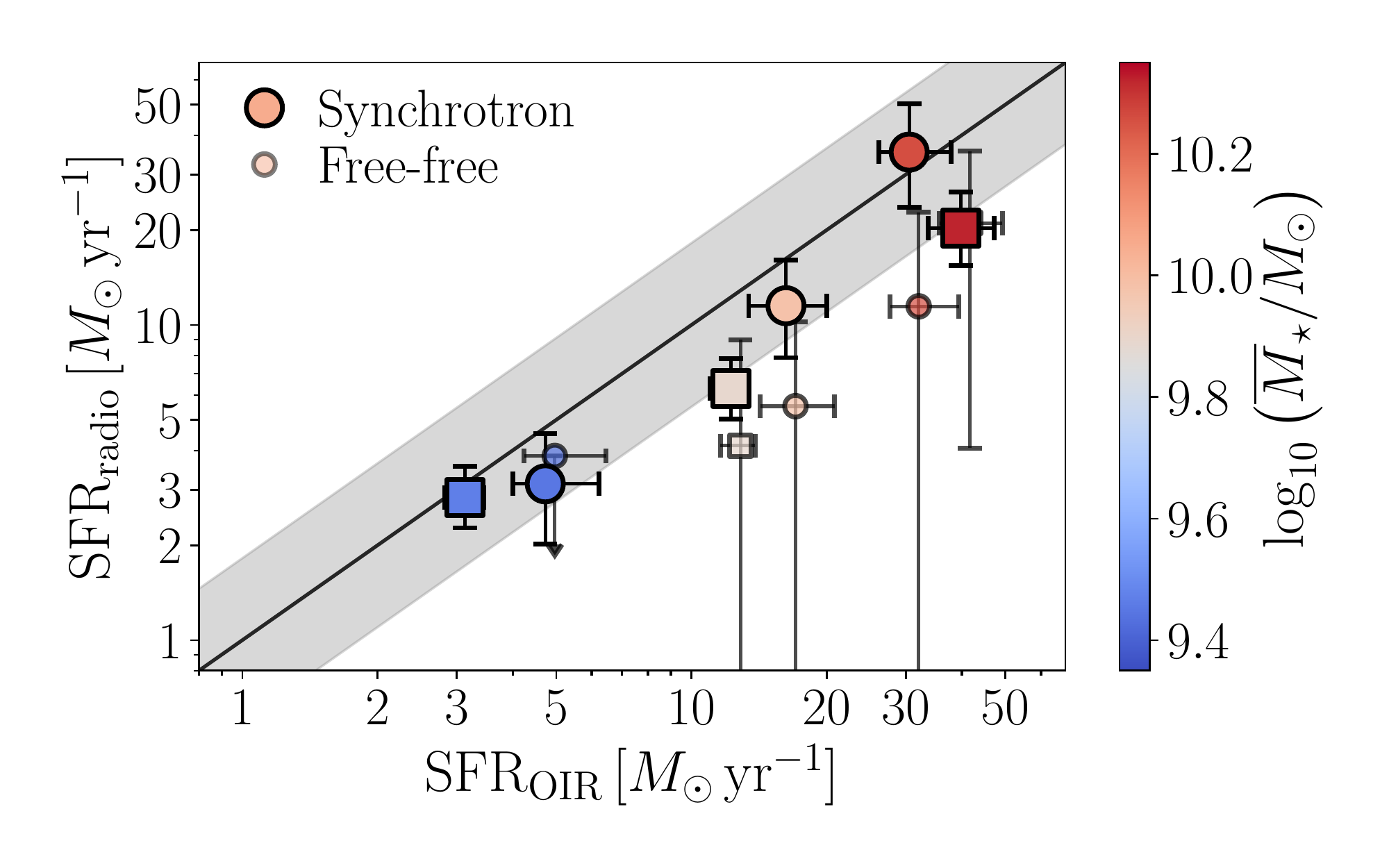}
     \includegraphics[width=0.49\textwidth]{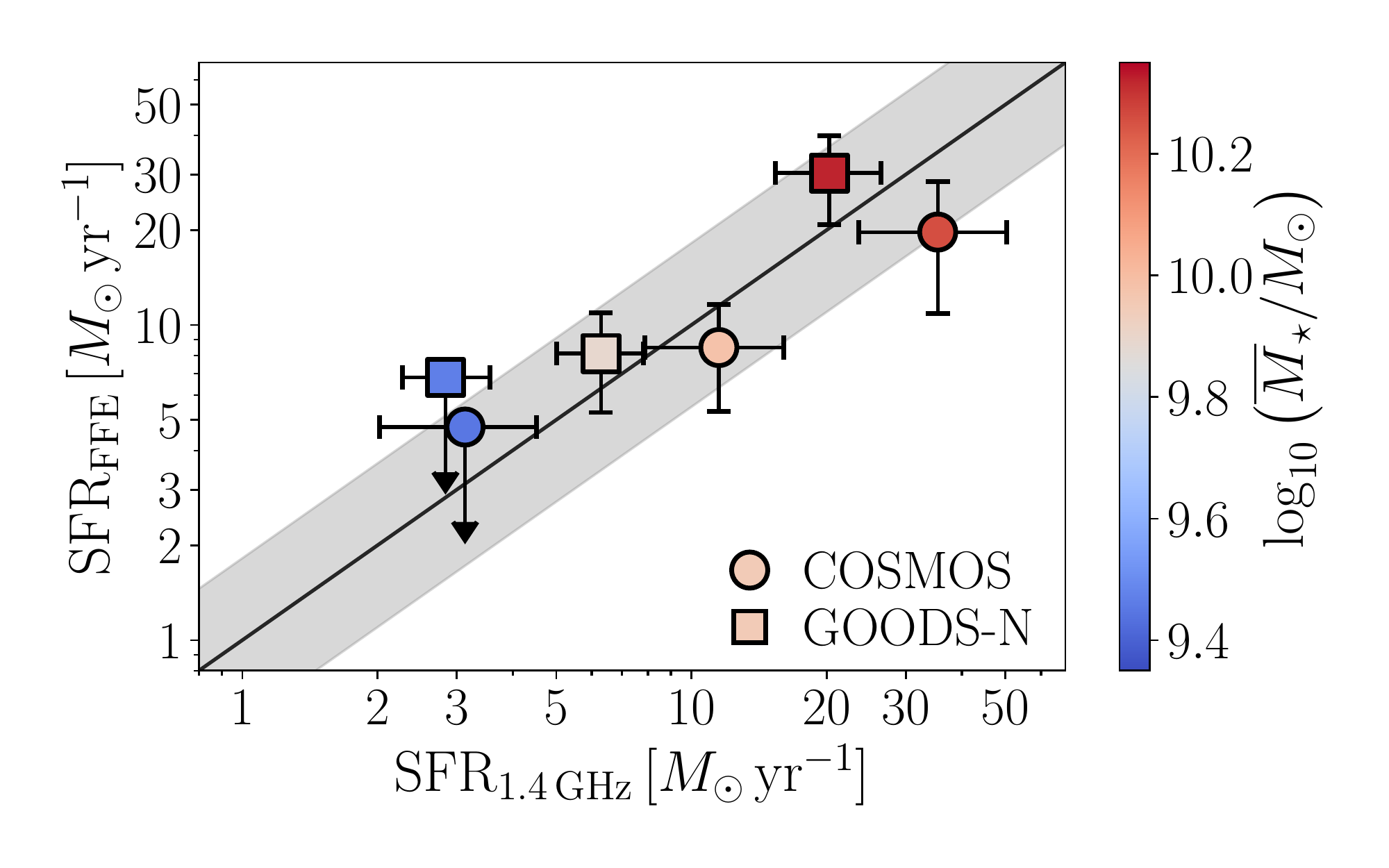}
     \vspace*{-0.3cm}
    \caption{\textbf{Left:} Comparison of the radio-based star formation rates and those from SED-fitting in the COSMOS (circles) and GOODS-N (squares) fields. The grey shading indicates the scatter about the local far-infrared/radio correlation from \citet{bell2003}. While the synchrotron star formation rates are in reasonable agreement with the optical-IR SFRs, the free-free SFRs are significantly lower when the fitted thermal fractions are adopted. \textbf{Right:} Comparison of the synchrotron and free-free star formation rates when an M82-like SED (thermal fraction of $f_\text{th}(34\,\text{GHz}) \approx 0.7$) is assumed. Given this thermal fraction, the radio-based SFRs are in excellent agreement.}
    \label{fig:stacked_sfrs}
\end{figure*}

We proceed by determining synchrotron and free-free star formation rates from the stacked radio luminosities, starting with the former. In each of the mass-complete bins, we have a clear stacked detection at 1.4 and 3 or 5\,GHz. We adopt the corresponding spectral index $\alpha^{1.4}_{3/5}$ in order to calculate the $K$-corrected luminosity density at rest-frame 1.4\,GHz, which is the conventional normalization frequency of the far-infrared/radio correlation. As in Section \ref{sec:results_optical}, we then adopt the mass-dependent parameterization of the far-infrared/radio correlation from \citet{delvecchio2021} and calculate low-frequency radio star formation rates via Equation \ref{eq:synchrotron_sfr}. We compare these star formation rates with those determined from OIR SED fitting in the left panel of Figure \ref{fig:stacked_sfrs}. The COSMOS and GOODS-N galaxy samples span a similar range of star formation rates, ranging from an average $\text{SFR}_\mathrm{1.4\,GHz} \approx 3 - 30 \,M_\odot\,\text{yr}^{-1}$ in the low and high mass bins, respectively. The synchrotron SFRs correlate well with the OIR SFRs, supporting the robustness of our stacking analysis. However, we find the synchrotron SFRs to be slightly lower than the optical SFRs by an average of $-0.16\,$dex (scatter of $0.13\,$dex). This may be related to the uncertain nature of far-infrared/radio correlation in low-mass galaxies at high redshift, as these are typically not individually detected even in deep radio imaging. In addition, \citet{driver2018} highlight minor but systematic differences between the star formation rates derived across various SED fitting codes. They show that both 3D-HST and, in particular, COSMOS2015 predict higher star formation rates than {\sc{magphys}} \citep{dacunha2008} in the range of interest for our analysis, with a difference of up to $\sim0.2\,$dex (see also \citealt{dudzeviciute2020,leja2021,thorne2021}). In turn, if the optical/IR star formation rates are indeed overestimated, this would reconcile the offset seen between these and the radio SFRs. Given both the uncertain nature of the far-infrared/radio correlation and systematic uncertainties in star formation rates from SED fitting, we conclude that the radio and OIR SFRs are in reasonable agreement.


We additionally present the free-free star-formation rates, adopting the fitted thermal fractions (Figure \ref{fig:stacked_sfrs}; left panel). As expected from the low thermal fractions, the free-free SFRs tend to be significantly lower than the optical-infrared SFRs, although the individual uncertainties on the former are large. We discuss this apparent deficit of high-frequency radio emission in Section \ref{sec:discussion_steepening}. However, it is interesting to additionally consider the free-free star formation rates we would infer when assuming a simple default value for the thermal fraction, as one may do when no multi-frequency radio data are available for a spectral decomposition. Upon adopting an M82-like starburst model ($f_\text{th} \sim 0.7$) we find good agreement between the synchrotron and free-free star-formation rates (Figure \ref{fig:stacked_sfrs}; right panel). This, in turn, indicates that the true thermal fraction exceeds the values derived from the spectral decomposition.

While free-free and synchrotron emission trace star formation on different timescales, our stacking analysis ensures we average across the star-formation histories of our galaxy sample. In turn, we can assume the free-free and synchrotron star formation rates equal one another, and use this to determine the thermal fraction via $\text{SFR}_\text{FFE}\left(f_\text{th} = 1\right) = f_\text{th} \times \text{SFR}_\text{1.4\,GHz}$. A linear fit to the SFRs results in a thermal fraction of $f_\text{th} = 0.77_{-0.18}^{+0.25}$ at observed-frame 34\,GHz. This is consistent with the expected thermal fraction from an M82-like radio spectrum at $z=1$, which predicts $f_\text{th} \approx 0.70$ at observed-frame 34\,GHz. While we probe slightly different rest-frame frequencies across the bins (Table \ref{tab:binning}), the variation in the thermal fraction of an M82-like SED between $z=0.9$ and $z=1.6$ -- the typical redshift of bins $M$-low and $M$-high -- is only $\Delta f_\text{th} \approx 0.05$, well within the errors of our fitted thermal fraction.

While our fitting routine prefers relatively low thermal fractions at 34\,GHz, resulting in low free-free SFRs, the above analysis indicates that a typical thermal fraction of $f_\text{th} \sim 0.7 - 0.8$ produces SFRs that are in better agreement with SFRs derived from synchrotron emission and optical-infrared SED-fitting. We discuss this finding in detail in the following Section.

\subsection{A Lack of High-frequency Emission}
\label{sec:discussion_steepening}

The apparent lack of high-frequency radio emission can arise in two possible ways. It is possible that the faint, star-forming population may be deficient in high-frequency free-free emission. Alternatively, this population may lack high-frequency synchrotron emission, which is indicative of a more complex radio spectrum than the canonical M82 starburst model. In what follows, we discuss both of these possibilities.

\subsubsection{A Lack of Free-Free Emission}

First, we discuss our findings in light of a deficit of high-frequency free-free emission. The simplest explanation of such a deficit requires a non-negligible optical depth of free-free emission, $\tau_\nu^\text{FF}$, at the frequencies probed. However, given the strong frequency dependence of $\tau_\nu^\text{FF} \propto \nu^{-2.1}$, free-free emission is certainly optically thin at observed-frame 34\,GHz. Alternatively, the low radio frequencies could be affected by free-free absorption. In this regime, the radio spectrum is dominated by synchrotron emission, and hence any significant free-free absorption should give rise to shallower synchrotron spectra. This, in turn, may cause the high-frequency radio spectrum, where free-free absorption does not play a role, to be steeper relative to the low frequencies where the optical depth is not negligible, throwing off the spectral fitting. 

This interpretation, however, appears unlikely as we find typical synchrotron slopes ($\alpha_\text{NT} \sim -0.85$) for our stacks, while any free-free absorption should flatten this value. In addition, optical depth effects are generally limited to $\nu \ll 1\,$GHz for modestly star-forming galaxies \citep{condon1992}, as probed in this work. Even for brighter star-forming galaxies, such as the local (U)LIRGs studied by \citet{murphy2013}, the typical frequency at which the spectrum turns over is $\nu \sim 1\,$GHz. In addition, bright $z\sim2$ submillimeter-detected starbursts show typical radio spectra of $\alpha\sim-0.80$ between observed-frame frequencies of $610\,\text{MHz} - 1.4\,\text{GHz}$ (roughly probing rest-frame $2 - 5\,$GHz; \citealt{ibar2010,thomson2014,algera2020a}), and hence do not show any evidence for spectral flattening due to free-free absorption. Using sensitive 150\,MHz observations of high-redshift starbursts, \citet{ramasawmy2021} further find that free-free absorption is typically limited to rest-frame frequencies $\nu\lesssim1\,$GHz. Given that we probe more modestly star-forming galaxies in this work (average $\text{SFR}_\text{1.4\,GHz} \approx 3 - 30\,M_\odot\,\text{yr}^{-1}$; Figure \ref{fig:stacked_sfrs}) at rest-frame frequencies $\nu \geq 2\,$GHz, we conclude that free-free absorption is unlikely to significantly affect the frequencies sampled in this work, and as such the low-frequency radio spectra should be well-described by a combination of power-law free-free and synchrotron emission.

The low fitted thermal fractions could alternatively point towards a synchrotron excess in galaxies. For example, \citet{murphy2013} determine a typical thermal fraction of $f_\text{th}(1.4\,\text{GHz}) \approx 0.05$ for a sample of 31 local ULIRGs, which is lower than the canonical M82-like value. They interpret this through dynamical effects, whereby merging systems form synchrotron bridges between the individual galaxies (see also \citealt{condon1993}). While not reducing a galaxy's free-free luminosity, such a synchrotron excess naturally results in lower thermal fractions. However, while the local ULIRG population tends to be dominated by merging systems (e.g., \citealt{armus1987}), less than 10\% of galaxies on the $z \lesssim 2$ star formation main sequence appear to be major mergers (e.g., \citealt{lopez-sanjuan2009,ventou2017,cibinel2019}; though see \citealt{puglisi2019}). In addition, a comparison of synchrotron star formation rates with those from SED-fitting (Section \ref{sec:discussion_sfrs}) does not show any evidence for a synchrotron excess. As such, we disfavor the scenario whereby the low thermal fractions are the result of an excess in synchrotron emission.

Alternatively, a deficit of free-free emission may arise when Lyman continuum photons are absorbed by dust still within the star-forming regions \citep{inoue2001,dopita2003}, or instead when a significant fraction of ionizing photons leaks out of the regions (see also \citealt{querejeta2019}). The former scenario was indeed invoked by \citet{barcosmunoz2017} to explain the low thermal fractions observed in a sample of local ULIRGs. ULIRGs, however, are compact and strongly dust-obscured systems, while in local, more modestly star-forming galaxies, free-free emission is observed to correlate well with dust-corrected SFRs from H$\alpha$ and $24\,\mu$m emission \citep{tabatabaei2017}. At high redshift, \citet{murphy2017} determine typical thermal fractions based on $1.4 - 10\,$GHz spectral indices that are consistent with the expected level of free-free emission from an M82-like radio spectrum. Similarly, \citet{algera2020c} determine free-free star-formation rates for a 34\,GHz selected sample that are in good agreement with those from synchrotron emission and SED-fitting. Moreover, while free-free emission is only affected by dust attenuation within H\,{\normalsize II} regions, SFR tracers using ionized gas at shorter wavelengths, such as the Balmer lines, should be affected by dust attenuation throughout the entire galaxy. However, with the possible exception of highly dust-obscured starbursts \citep{chen2020}, the Balmer lines have been shown to agree both with panchromatic SFRs derived via SED-fitting \citep{shivaei2016} and SFRs from radio synchrotron emission \citep{duncan2020} at $z\sim2$. Given that free-free emission is expected to be less affected by dust than H$\alpha$, it seems unlikely that high-redshift galaxies with modest star-formation rates exhibit a systematic deficit of free-free emission.

\subsubsection{A Lack of Synchrotron Emission}

\begin{figure*}[!t]
    \centering
    \vspace*{-1.5cm}
    \includegraphics[width=0.75\textwidth]{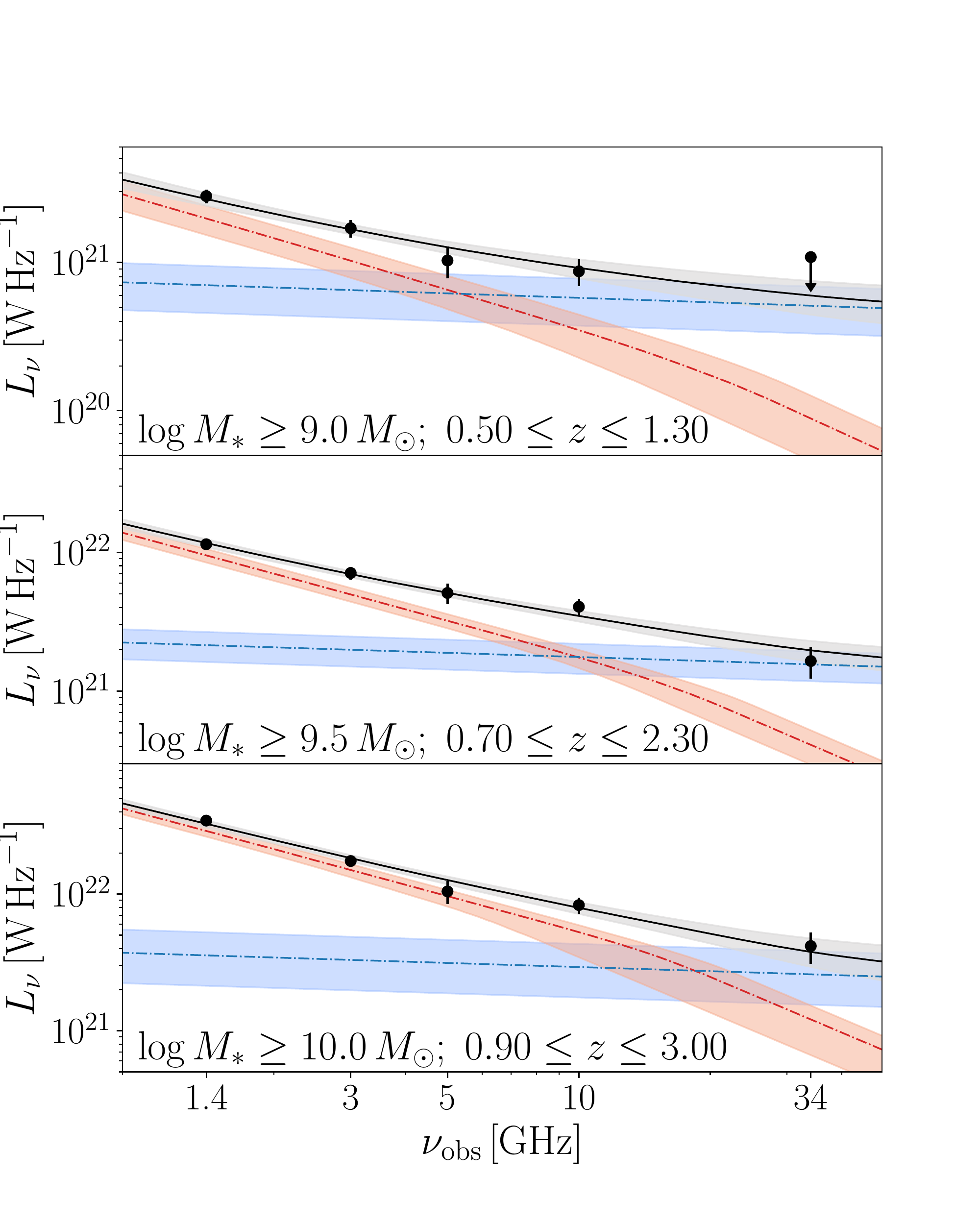}
    \vspace*{-0.8cm}
    \caption{Stacked radio spectra combining the photometry across the COSMOS and GOODS-N fields (see Figures \ref{fig:stacked_cosmos} and \ref{fig:stacked_goodsn}) to sample the radio spectrum at five distinct frequencies. This additional constraint enables fitting the combined spectrum (grey) with a model of free-free emission (blue) and synchrotron emission including a break (red). In contrast to the fits that do not incorporate a break, we now find that the observed-frame 34\,GHz luminosity is likely dominated by free-free emission.}
    \label{fig:combined_spectra}
\end{figure*}

The observed lack of high-frequency emission may instead be due to a deficit of synchrotron emission at 34\,GHz. This is supported by the fact that in all stacks the observed-frame 34\,GHz luminosities are in good agreement with the free-free luminosities predicted based on optical-infrared star formation rates. This, therefore, is indicative of a large thermal contribution, and hence a lack of synchrotron emission. A high-frequency deficit of synchrotron emission is most readily interpreted as synchrotron aging: high-energy cosmic rays, which emit predominantly at high frequencies, are the first to radiate their energy via synchrotron emission. Such synchrotron cooling has been invoked to explain steep synchrotron spectra in local spiral galaxies \citep{tabatabaei2017}, as well as spectral steepening in bright starbursts, both locally (e.g., \citealt{colbert1994,clemens2008}) and at high redshift \citep{thomson2019}.

\begin{figure*}[!t]
    \centering
    \vspace*{-2.0cm}
    \includegraphics[width=0.75\textwidth]{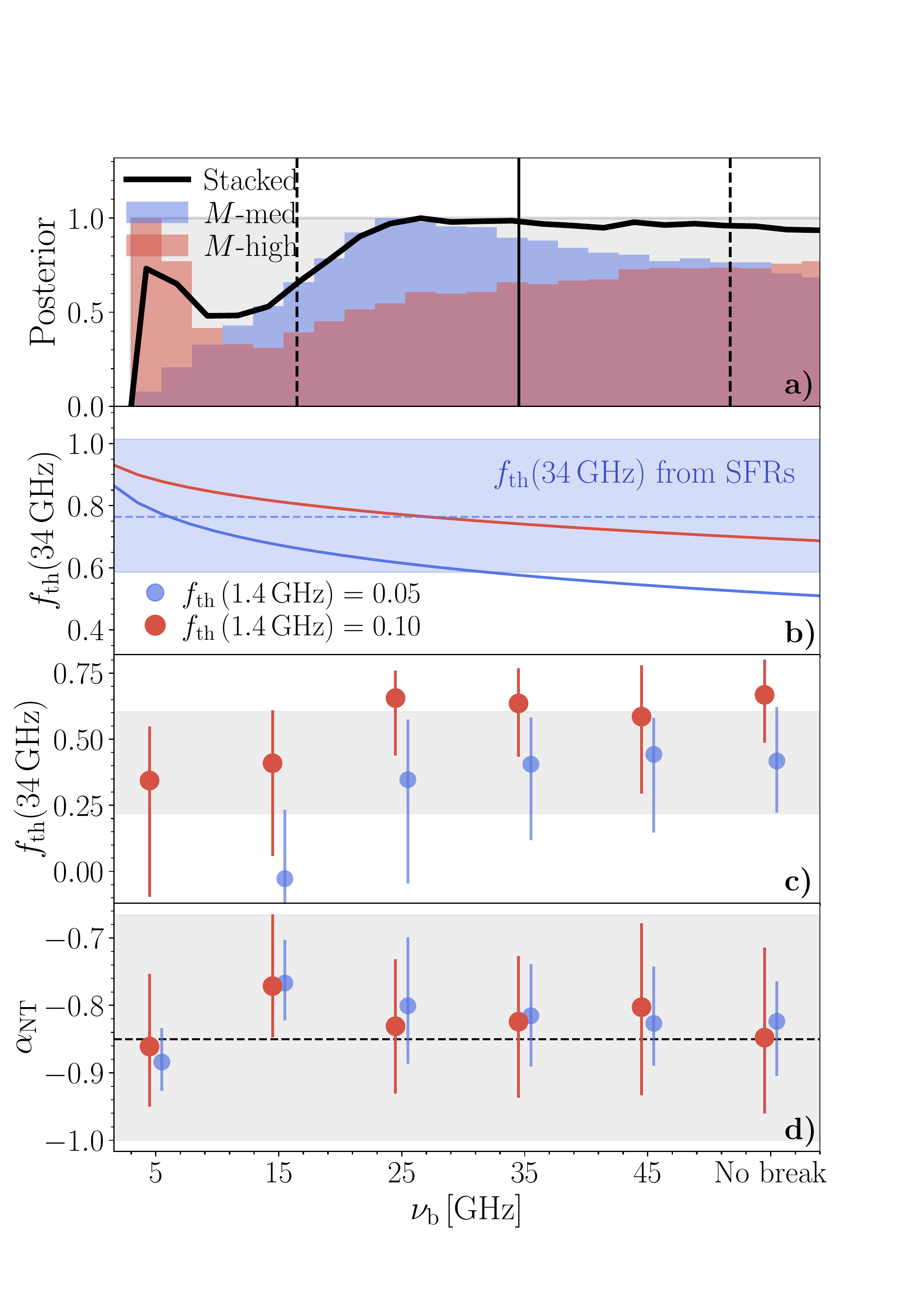}
    \vspace*{-1.3cm}
    \caption{A quantitative test of synchrotron aging in the radio spectra of typical high-redshift star-forming galaxies. \textbf{a)} Posterior distributions of the break frequency $\nu_b'$ in bins $M$-med and $M$-high, obtained after explicitly including a spectral break in the fitting. The combined posterior is shown through the solid black line, while the prior is shown through the grey rectangle. The solid and dashed vertical lines indicate the median and $16-84^\text{th}$ percentiles from the combined posterior, respectively. High break frequencies $\nu_b' \gtrsim 15\,$GHz are favored by the fitting routine. \textbf{b)} The 34\,GHz thermal fraction as a function of $\nu_b'$ for an M82-like spectrum at $z=1$, given a 1.4\,GHz thermal fraction of 10\% (as in M82; red line) and 5\% (as in \citealt{murphy2013}; blue line). The shaded blue region shows the expected thermal fraction based on the stacked synchrotron SFRs. A 5\% thermal fraction at 1.4\,GHz requires the existence of a low-frequency break ($\nu_b' \lesssim 25\,$GHz) in order to match the measured $f_\text{th}(34\,\text{GHz})$. This, however, is disfavored by the posterior in panel a). \textbf{c) \& d)} Illustration of the biases induced when fitting complex spectra with a simple model. The recovered thermal fractions (c) and synchrotron spectral indices (d), obtained from fitting simulated radio spectra, are shown as a function of break frequency. The colored points indicate the median recovered value among the simulations, given an input thermal fraction at rest-frame 1.4\,GHz, and the 16-84 percentile spread. The grey bands indicate the observed range of 34\,GHz thermal fractions and synchrotron slopes when fitting the combined COSMOS and GOODS-N stacks with our model in Equation \ref{eq:spectrum}. A break in the radio spectrum is plausible when the observed and simulated spread among both $f_\text{th}$ and $\alpha_\text{NT}$ are similar, though the latter parameter does not provide meaningful constraints. A combination of either $f_\text{th}(1.4\,\text{GHz}) = 0.05$ and $\nu_b' \gtrsim 30\,$GHz or $f_\text{th}(1.4\,\text{GHz}) = 0.10$ and $15\,\text{GHz} \lesssim \nu_b' \lesssim 25\,$GHz provides a reasonable match between simulations and observations. Upon combining the constraints across the four panels, we qualitatively infer that the radio spectrum of star-forming galaxies can be characterized by a typical thermal fraction ($f_\text{th}(1.4\,\text{GHz}) \sim 0.10$) and a spectral
    break at $\nu_b' \sim 15 - 25\,$GHz.}
    \label{fig:mockfitting}
\end{figure*}

As this work investigates the radio spectra of faint star-forming galaxies, we are limited by both the signal-to-noise ratio at high frequencies, and by the sampling of the radio SED at four distinct frequencies across both the COSMOS and GOODS-N fields. This, in turn, makes it difficult to fit a more complex prescription of the radio spectrum to the available photometry. However, we may improve the sampling of our spectra by combining the available radio data across COSMOS and GOODS-N, as both fields have only three out of four frequencies in common. Given that we adopt identical mass-complete bins in both fields, we expect to trace the same underlying galaxy population, which is supported by the measured radio luminosities across both fields being in good agreement (Table \ref{tab:stacking}).

In order to maximize the S/N across the combined spectrum, we re-stack in the GOODS-N field at 1.4, 5 and 34\,GHz. By not including the 10\,GHz observations, we are able to use the entire 34\,GHz footprint in GOODS-N for stacking, while we adopt the deeper 10\,GHz constraints from COSMOS. The re-stacked luminosities in GOODS-N are in agreement with those provided in Table \ref{tab:stacking}, though attain a slightly higher S/N. We subsequently construct an average radio spectrum by taking a noise-weighted mean between the 1.4\,GHz and 34\,GHz observations across both COSMOS and GOODS-N, while adopting the 3 and 10\,GHz stacks from our COSMOS analysis and the 5\,GHz stacks in GOODS-N. As such, combining the two fields amalgamates radio continuum data across five frequencies within $1.4 - 34\,$GHz. We have verified that excluding the GOODS-N 10\,GHz data does not affect our conclusions in the following sections.

We first re-fit the combined radio spectra with our simple model of free-free and synchrotron emission (Equation \ref{eq:spectrum}). In the two bins with a stacked detection at 34\,GHz, $M$-med and $M$-high, we determine thermal fractions of $f_\text{th}(34\,\text{GHz}) = 0.23_{-0.63}^{+0.35}$ and $0.52_{-0.28}^{+0.18}$, respectively. These values, while uncertain, are consistent with the thermal fractions determined for the COSMOS and GOODS-N fields individually, and are hence lower than expected from an M82-like spectrum. However, as the combined spectra span five frequencies, we can attempt to fit a more complex model to the available photometry, incorporating spectral aging. We adopt the standard synchrotron aging model whereby the non-thermal spectral index steepens to $\alpha_\text{NT} - 0.5$ beyond a break frequency $\nu_b$ \citep{kardashev1962}. We note that this model is a simplification, and assumes the galaxy star formation histories may be characterized by a single burst. Nevertheless, we note that more complicated continuous star-formation histories induce similar spectral behavior \citep{thomson2019,algera2020a}. The resulting functional form of the radio spectrum may be written as

\small
\begin{align}
    L_{\nu'} = 
    \begin{cases} 
          L_{\nu_0'} \left[ \left( 1-f_{\nu_0'}^\text{th} \right) \left( \frac{\nu'}{\nu_0'}\right)^{\alpha_\text{NT}} + f_{\nu_0'}^\text{th} \left( \frac{\nu'}{\nu_0'}\right)^{-0.1} \right]
          & \nu' \leq \nu_b' \\
          L_{\nu_b'} \left[ \left( 1-f_{\nu_b'}^\text{th} \right) \left( \frac{\nu'}{\nu_b'}\right)^{\alpha_\text{NT} - 0.5} + f_{\nu_b'}^\text{th} \left( \frac{\nu'}{\nu_b'}\right)^{-0.1} \right]
          & \nu' > \nu_b'
    \end{cases}
    \label{eq:spectrum_break}
\end{align}
\normalsize

\noindent where the various parameters have the same meaning as in Equation \ref{eq:spectrum}, and $\nu_b' > \nu_0'$ is assumed. The modest S/N of our high-frequency data does not allow us to freely vary all parameters in the fitting. In what follows, we therefore adopt a fixed $\alpha_\text{NT} = -0.85$. As before, we assume a flat prior on the thermal fraction and overall normalization, and in addition we now adopt a flat prior on the break frequency of $\nu_b'\in[3,60]\,$GHz. This range was adopted to constrain the break frequency within the typical range of rest-frame frequencies sampled by our stacks at $z\sim0.9-1.6$ (Table \ref{tab:radio}). We show the radio stacks combining the COSMOS and GOODS-N fields and their decomposition into free-free and (steepened) synchrotron emission in Figure \ref{fig:combined_spectra}.

In the bins with a stacked 34\,GHz detection, $M$-med and $M$-high, we find typical break frequencies of, respectively, $\nu_b' = 34_{-15}^{+17}\,$GHz and $\nu_b' = 35_{-24}^{+17}\,$GHz. In addition, we recover thermal fractions of $f_\text{th}(1.4\,\text{GHz}) = 0.11 \pm 0.03$ and $f_\text{th}(1.4\,\text{GHz}) = 0.06_{-0.02}^{+0.03}$ for the two bins, in agreement with the canonically assumed thermal fraction of M82 of 10\%. However, given that we place a flat prior on the break frequency, we would expect that when the break does not significantly affect the fitting, we recover our prior distribution, and hence a fitted spectral break around the average frequency of $\nu_b' \sim 30\,$GHz. To investigate whether this is indeed the case, we show the individual and combined posterior distributions on the break frequency in bins $M$-med and $M$-high in panel a) of Figure \ref{fig:mockfitting}. A break at low frequencies ($\nu_b' \lesssim 15\,$GHz) is disfavored, while at higher frequencies the posterior distribution is flat, indicating a wide range of plausible break frequencies. As such, while we cannot precisely determine the location of a spectral break, we may conclude that if a break exists, it is likely to arise at a frequency $\nu_b' \gtrsim 15\,$GHz. 

Next, we turn to our constraints on the 34\,GHz thermal fraction based on a comparison of synchrotron and free-free star formation rates. Under the assumption that the free-free and synchrotron SFRs are identical, we determined a thermal fraction of $f_\text{th}(34\,\text{GHz}) = 0.77_{-0.18}^{+0.25}$ (Section \ref{sec:discussion_sfrs}). We compare this value with the expected high-frequency thermal fraction as a function of break frequency in panel b) of Figure \ref{fig:mockfitting}. For an M82-like radio spectrum at $z=1$ with $f_\text{th}=0.10$ at 1.4\,GHz, we expect a 34\,GHz thermal fraction consistent with the predicted value for any given frequency of a spectral break. However, for a low thermal fraction of $f_\text{th}(1.4\,\text{GHz}) = 0.05$, a low-frequency break at $\nu_b'\lesssim25\,$GHz is required to match the predicted high-frequency thermal fraction. This constraint is therefore rather orthogonal to the constraints on $\nu_b'$ from the posterior distributions in Figure \ref{fig:mockfitting}a.

Finally, we investigate whether the existence of a spectral break can indeed result in low thermal fractions when unaccounted for in the fitting. To this end, we simulate the radio spectra of faint star-forming sources with a spectral break at frequency $\nu_b'$. We then fit the resulting simulated radio spectrum with a simple combination of free-free and synchrotron emission, as in Equation \ref{eq:spectrum}. This model by construction cannot capture any spectral aging, and as such simulates how any complexities in the radio spectra of star-forming galaxies might leave their imprint when modelled by our simple fitting routine. For the radio spectra we adopt an M82-like model, but with a synchrotron component that steepens beyond $\nu_b$, and fix the S/N at 34\,GHz to be similar to what is observed in our stacks ($\text{S/N}\approx3 - 5$). Using this model for the radio spectrum, we further sample the fluxes of the simulated spectra at 1.4, 3, 5 and 10\,GHz, and vary all fluxes within their corresponding uncertainties. We subsequently fit the simulated radio fluxes with our simple model a total of 400 times each for a range of assumed break frequencies spanning $\nu_b' = 5 - 45\,$GHz. 

We show the recovered thermal fractions and synchrotron spectral indices as a function of break frequency for the simulated sources in panels c) and d) of Figure \ref{fig:mockfitting}. We can naturally reproduce our low recovered thermal fractions by using an M82-like radio SED without any spectral break, provided that the input thermal fraction is low ($f_\text{th} \sim 0.05$ at 1.4\,GHz). In this case, a thermal fraction of $f_\text{th} \sim 0.50$ at 34\,GHz is expected. However, as is evident from panel b), this underestimates the predicted thermal fractions based on a comparison of the stacked free-free and synchrotron SFRs. Instead, a larger thermal fraction at rest-frame 1.4\,GHz, in combination with a spectral break, can similarly give rise to modest fitted thermal fractions, while the true thermal fraction is significantly higher. Indeed, these simulations indicate that a likely value for the thermal fraction is $f_\text{th}(1.4\,\text{GHz}) > 0.05$, with a combination of $f_\text{th}(1.4\,\text{GHz}) = 0.10$ and $\nu_b' \approx 15 - 25\,$GHz being able to explain the observed spectral parameters. \\

Summarizing, we have tested whether the presence of a break in the synchrotron spectrum can plausibly explain the low fitted thermal fractions. Directly constraining the location of the spectral break via fitting a more complex radio spectrum to the available photometry indicates $\nu_b'\gtrsim15\,$GHz. A comparison of the synchrotron and free-free SFRs points towards either a low thermal fraction and a low-frequency spectral break, or an overall high thermal fraction. Simulations that involve fitting radio spectra with a break using a model that does not account for spectral aging indicates a high thermal fraction and a break at $\nu_b'\approx 15 - 25\,$GHz. Upon combining these constraints, we qualitatively conclude that the radio spectrum of typical star-forming galaxies may be described with a typical low-frequency thermal fraction ($f_\text{th} \approx 0.10$), and a spectral break at moderately high-frequencies ($\nu_b'\approx 15 - 25\,$GHz). If indeed the result of spectral aging, such a break reflects a population of older, cooled cosmic ray electrons and hence provides a rough proxy of a galaxy's recent star formation history (e.g., \citealt{thomson2019}).

However, we caution that while this interpretation qualitatively explains the observed deficit of high-frequency radio emission, at present we cannot test it in detail. Instead, testing this hypothesis will require an increased sampling of the radio SED at high frequencies, in particular around observed-frame $20\,$GHz, allowing for more complex models for the radio spectrum to be fit to the data. At present, such analyses have remained limited to relatively nearby starburst galaxies (e.g., \citealt{galvin2018}). However, extending such studies to higher redshifts will be crucial in order to establish free-free emission as an SFR indicator in the early Universe.

In addition, we note that our analysis is limited to relatively massive systems (typical mass of $\log \left(M_\star/M_\odot\right) \sim 10$; Table \ref{tab:binning}), owing to the lack of a 34\,GHz detection in our low-mass bin. Furthermore, given the nature of stacking involves averaging across large galaxy samples, the present analysis cannot conclusively ascertain whether a potential spectral break is common throughout the high-redshift star-forming galaxy population, or is more pronounced for any particular subset. Given that a spectral break typically arises under the influence of strong magnetic fields ($\nu_b \propto B^{-3}$ at fixed galaxy age; \citealt{carilli1996}), and magnetic fields may be enhanced by high star formation surface densities (e.g., $B \propto \Sigma_\mathrm{SFR}^{1/3}$; \citealt{schleicher2013}), extreme starbursts may be most affected by spectral aging. Assuming a linear slope for the galaxy main sequence (e.g., \citealt{speagle2014}), as well as an approximately flat relation between stellar mass and radio size \citep{jimenez-andrade2019}, we may infer that, roughly, $B \propto M_\star^{1/3}$ and hence $\nu_b \propto M_\star^{-1}$. This, in turn, implies that synchrotron aging is expected to be most pronounced for massive galaxies, in qualitative agreement with the recent results from \citet{an2021} at lower frequencies. However, more data are clearly needed to test this scenario in detail.

\subsection{The Cosmic Star-formation History}

In Section \ref{sec:results_optical}, we detect multi-frequency radio continuum emission in star-forming galaxies in COSMOS and GOODS-N, based on a median stacking analysis. We now perform a mean stacking analysis in order to measure the average radio luminosity of star-forming galaxies at 34\,GHz, and in turn compute their corresponding average star formation rate. Given that we are stacking on mass-complete galaxy samples within a known cosmic volume, we can directly constrain the cosmic star-formation rate density with free-free emission.

However, as we discussed in Section \ref{sec:discussion_steepening}, the fitted thermal fractions may be underestimated if synchrotron aging affects the radio spectrum, which in turn results in underestimated cosmic star-formation rates. As such, in what follows, we calculate star formation rates in three different ways, assuming either 1) the fitted thermal fractions combining the COSMOS and GOODS-N fields from Section \ref{sec:discussion_steepening}; 2) a fixed thermal fraction of unity or; 3) the thermal fraction obtained from comparing the free-free and synchrotron SFRs in Section \ref{sec:discussion_sfrs}. In the first case, we are likely underestimating the true cosmic SFRD, while the measurements from case three are not fully independent of the synchrotron-derived cosmic SFRD. In turn, the most direct and unbiased constraints on the SFRD from radio free-free emission are provided by case two, which constitute an upper limit on the total rate of cosmic star formation.

\begin{figure}[!t]
    \centering
    \hspace*{-1.2cm}\includegraphics[width=0.6\textwidth]{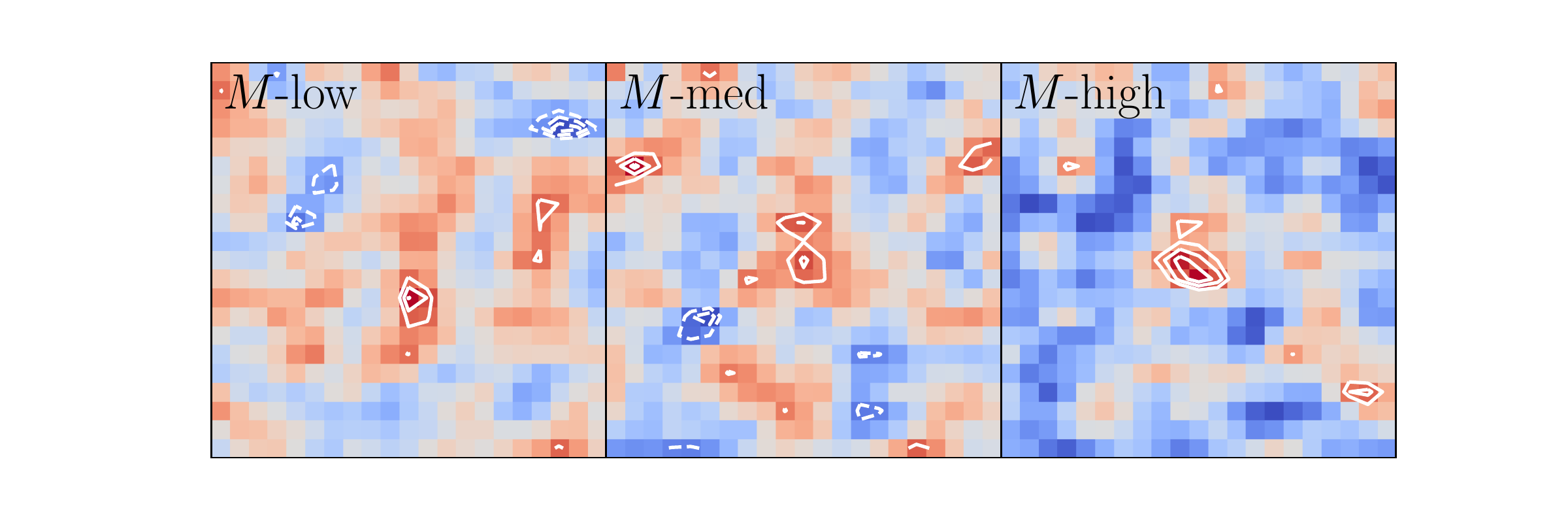}
    \vspace*{-0.5cm}
    \caption{Mean stacks at 34\,GHz for the mass-complete bins combining the COSMOS and GOODS-N fields. Only the central $21\times21$ pixels are shown for clarity, and the contours represent $2\sigma$ to $4\sigma$ in steps of $0.5\sigma$. Continuum emission in the stacks is detected at $3.0\sigma$ and $3.7\sigma$ in the medium and high-mass bins, while no significant emission in $M$-low is observed at the $2.5\sigma$ level.}
    \label{fig:mean_stacks}
\end{figure}

\begin{figure*}[!t]
    \centering
    \includegraphics[width=\textwidth]{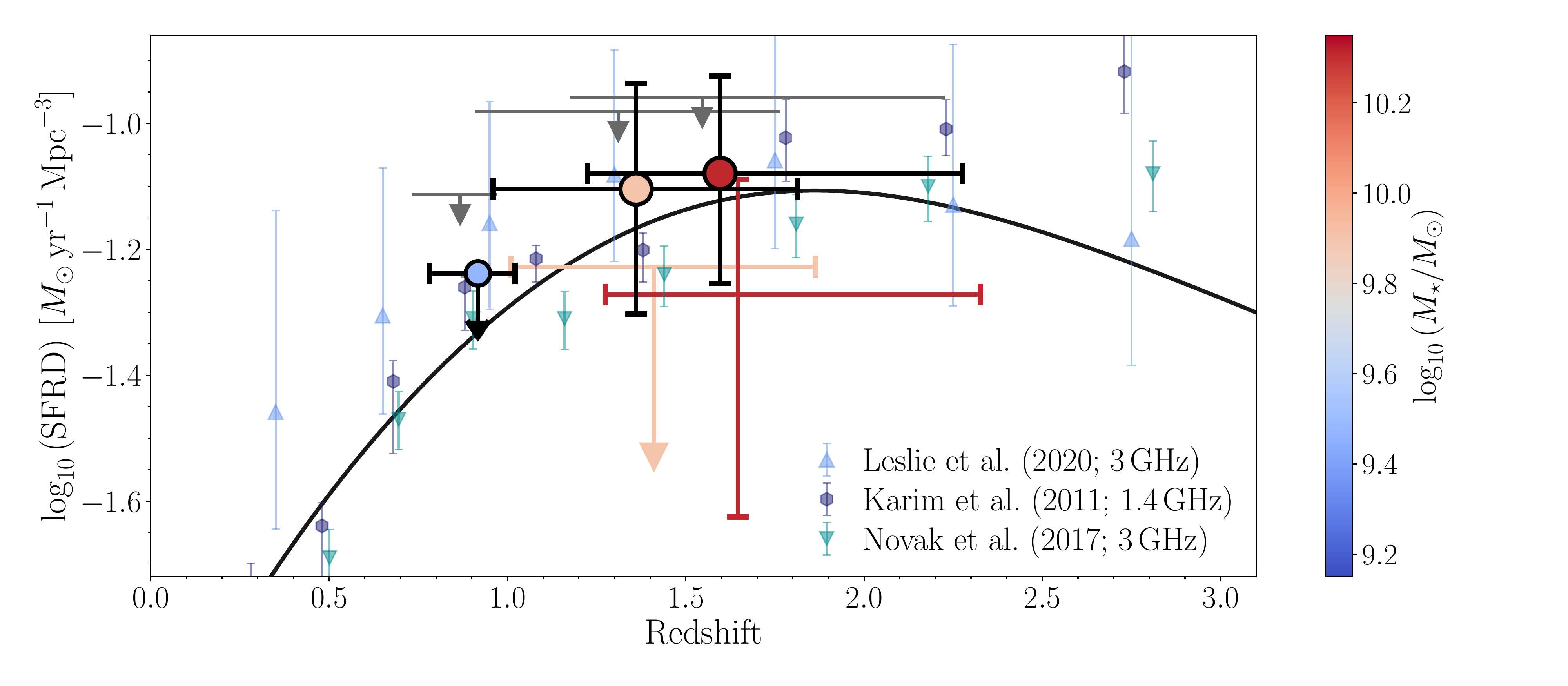}
    \caption{Constraints on the cosmic star formation history from free-free emission. The large points, placed at the median redshift of the corresponding bin, present the most robust constraints, and adopt a thermal fraction of $f_\text{th}(34\,\text{GHz}) = 0.77_{-0.18}^{+0.25}$, as determined through a comparison of synchrotron and free-free SFRs (Case 3; Section \ref{sec:discussion_sfrs}). Upper limits, assuming a thermal fraction of unity, are shown via grey arrows (Case 2). The colored errorbars show the cosmic star formation rate density when the fitted thermal fractions are adopted for the two bins with a stacked detection at 34\,GHz (Case 1; fits do not include a spectral break). The horizontal errorbars show the interquartile ranges for the corresponding redshift bins.  The canonical \citet{madau2014} cosmic SFRD is shown through the black curve, and various radio-based studies, using non-thermal synchrotron emission as a star-formation tracer, are additionally overplotted. Similar to this work, \citet{karim2011} and \citet{leslie2020} utilize a radio stacking analysis at low frequencies (1.4 and 3\,GHz, respectively), while \citet{novak2017} use radio-detected sources. Our constraints are in good agreement with those from commonly adopted star-formation tracers at high redshift, and show that radio free-free emission can be used to constrain cosmic star formation.}
    \label{fig:SFRD}
\end{figure*}

We first perform a mean stacking analysis to obtain the average radio luminosity at 34\,GHz, using the same mass bins as originally defined in Table \ref{tab:binning}. Note that we adopt the full COLD$z$ footprint in GOODS-N, as in Section \ref{sec:discussion_steepening}. We combine the cutouts around sources in both COSMOS and GOODS-N into a single cube of $N_\text{COS} + N_\text{GN}$ sources, and take a simple unweighted mean across the entire sample.\footnote{We note that the COSMOS and GOODS-N radio maps have the same pixel scale of $0\farcs5$, such that the stacks may easily be co-added.} The resulting stacks are shown in Figure \ref{fig:mean_stacks}. As the two fields have a slightly different beam, we do not fit the stacked emission with a 2D Gaussian as we did when median stacking, but simply extract the peak luminosity density within a radius of one pixel ($0\farcs5)$ from the center of the stack. In the medium and high-mass bins, we detect mean-stacked 34\,GHz emission at a significance of $3.0\sigma$ and $3.7\sigma$, respectively, while in the low-mass bin we do not detect significant continuum emission and instead adopt a $2.5\sigma$ upper limit. We subsequently add in the galaxies individually detected at 34\,GHz via Equation \ref{eq:meanstacking}, corresponding to six and four sources in bins $M$-med and $M$-high, respectively (out of the seven star-forming galaxies detected in the COLD$z$ 34\,GHz observations; \citealt{algera2020c}). In $M$-low, we determine an upper limit on the mean 34\,GHz luminosity of $L_{34} < 1.2 \times 10^{21}\,\rm{W\,Hz}^{-1}$, while we measure stacked luminosities of $L_{34} = (4.9 \pm 1.5)\times10^{21}\,\rm{W\,Hz}^{-1}$ and $L_{34} = (13.1 \pm 3.3)\times10^{21}\,\rm{W\,Hz}^{-1}$ in bins $M$-med and $M$-high, respectively.

Given the fitted or fixed thermal fractions, we subsequently calculate free-free star-formation rates via Equation \ref{eq:ffsfr}. The cosmic star formation rate density $\text{SFRD}(z)$ is then determined via

\begin{align}
    \text{SFRD}(\overline{z}) = C(M_*) \times \langle \text{SFR} \rangle \times N_\text{stack} \times V(z)^{-1} \ ,
\end{align}

\noindent where $\langle \text{SFR} \rangle$ is the average free-free star-formation rate, $N_\text{stack}$ is the number of sources used for stacking, combining detections and non-detections, $V(z)$ is the total cosmic volume probed by the 34\,GHz mosaic spanning the full redshift range of the bin, and $C(M_\star)$ is a numerical factor that corrects for the fact that some star-formation occurs below the stellar masses where we assume to be complete. In order to determine this correction, we adopt the stellar mass functions for star-forming galaxies from \citet{davidzon2017} corresponding to the median redshift in the bin. We further assume a linear relation between stellar mass and star formation rate, as is appropriate for low-mass ($M \lesssim 10^{10.5}\,M_\odot$) galaxies on the star-formation main-sequence \citep{speagle2014,schreiber2015,leslie2020}. The correction factor equals $C \approx 1.07$ and $C \approx 1.41$ for mass completeness limits of $10^{9}\,M_\odot$ and $10^{10}\,M_\odot$, respectively, when integrating the stellar mass function down to $10^{5}\,M_\odot$ (following \citealt{karim2011}). We note that, while the stellar mass function does evolve with redshift, the characteristic mass and low-mass slope for star-forming galaxies do not vary significantly within the redshift range probed in this work \citep{davidzon2017}. Indeed, evaluating the correction factor in the redshift range $0.5 < z < 2.5$ changes the resulting cosmic star formation rate density within less than $15\%$ -- well within the uncertainties on the stacked 34\,GHz luminosities.

We plot our free-free constraints on the cosmic star formation history in Figure \ref{fig:SFRD}, and provide the numerical values in Table \ref{tab:cosmic_sfrd}. When adopting the fixed case 3) thermal fraction, our cosmic star-formation rates obtained via free-free emission are in good agreement with the canonical SFRD from \citet{madau2014}. This is consistent with the fact that the SFRD derived from low-frequency radio observations similarly agrees with the \citet{madau2014} relation (\citealt{novak2017,leslie2020}; see also below). In addition, adopting a thermal fraction of unity places robust upper limits on the SFRD which are fully consistent with the \citet{madau2014} constraints. As expected, adopting the fitted thermal fractions (case 1) provides cosmic SFRs that are consistently biased low compared to the canonical SFRD, further highlighting that a thermal fraction of $f_\text{th}(34\,\text{GHz}) \sim 0.8$ is a good assumption for star-forming galaxies at $z\sim1$.

We caution that the normalization of the star formation main-sequence increases with redshift (e.g., \citealt{speagle2014}), and in turn varies across our relatively wide redshift bins. Since we adopt a luminosity stacking technique, the stacks are therefore weighted towards the on average more luminous high-redshift population within each bin. As a result, the luminosity-weighted typical redshift within the bins exceeds the median redshift of the sample shown in Figure \ref{fig:SFRD}. Qualitatively, if the luminosity-weighted redshift is adopted instead, the agreement between the SFRD from free-free emission and the \citet{madau2014} relation improves further, as the typical redshift probed now more closely approaches the peak of cosmic star-formation at $z\sim2$. In practice, we cannot determine the luminosity-weighted median redshift as the individual radio luminosities of our input sample are -- by construction -- unknown. However, adopting the optical/IR SFRs as a proxy, we determine median $\mathrm{SFR}_\mathrm{OIR}$-weighted redshifts for bins $M$-low, $M$-med and $M$-high of $z_\mathrm{low} \approx 1.0$, $z_\mathrm{med}  \approx 1.6$ and $z_\mathrm{high}  \approx 2.0$, respectively (c.f., Table \ref{tab:binning}). In turn, our constraints on the SFRD shown in Figure \ref{fig:SFRD} shift towards higher redshift when a weighted median is adopted, further improving the agreement with the \citet{madau2014} relation.

We finally compare our constraints with radio-based studies of cosmic star formation, which use low-frequency radio synchrotron emission. Both \citet{karim2011} and \citet{leslie2020} utilize stacking analyses in the COSMOS field, at 1.4\,GHz and 3\,GHz, respectively, while \citet{novak2017} consider individually detected radio sources. These radio-based studies may hint towards a slight excess in the SFRD compared to \citet{madau2014} at $z\gtrsim 2.5$, although the uncertain conversion from radio synchrotron emission into star formation rates at high redshift complicates such interpretations. While free-free emission does not require adopting such a conversion, at present it remains difficult to probe beyond the peak of cosmic star formation even via stacking analyses. In the future, radio observations at lower frequencies (for example at 10\,GHz, probing $\nu \gtrsim 30\,$GHz at $z\gtrsim2$) may provide better constraints in this high redshift regime, owing to the increased source brightness and larger field of view. Such observations may subsequently be used to distinguish between a cosmic star formation rate that follows the high-redshift decline of the \citet{madau2014} relation, or remains relatively flat, as might be expected if the current high-redshift UV- and optical-based constraints underestimate dust corrections (e.g., \citealt{casey2018}, but see also the constraints on the SFRD from \citealt{bouwens2020} and \citealt{zavala2021}).

\begin{deluxetable*}{cccc}
\label{tab:cosmic_sfrd}

\tablecaption{Constraints on the cosmic star formation rate density from free-free emission}
\tablehead{
    \colhead{Bin} &
	\colhead{$\log\left[\text{SFRD}(f_\text{th,obs})\right]$} & 
	\colhead{$\log\left[\text{SFRD}(f_\text{th} = 1)\right]$} &
	\colhead{$\log\left[\text{SFRD}(f_\text{th} = 0.77_{-0.18}^{+0.25})\right]$}
}

\startdata
 & $[M_\odot\,\text{yr}^{-1}\,\text{Mpc}^{-3}]$ & $[M_\odot\,\text{yr}^{-1}\,\text{Mpc}^{-3}]$ & $[M_\odot\,\text{yr}^{-1}\,\text{Mpc}^{-3}]$ \\
\tableline
$M$-low & $-$ & $< -1.09$ & $< -1.21$ \\
$M$-med & $< -1.20$ & $-0.96_{-0.15}^{+0.11}$ & $-1.08_{-0.20}^{+0.17}$ \\
$M$-high & $-1.25_{-0.36}^{+0.18}$ & $-0.94_{-0.12}^{+0.10}$ & $-1.06_{-0.18}^{+0.16}$
\enddata

\tablecomments{(1) Bin matches that in Table \ref{tab:binning}; (2) cosmic star-formation rate density when the fitted thermal fraction is used. When the thermal fraction is consistent with zero a $1\sigma$ upper limit on $f_\text{th}$ is used; (3) cosmic SFRD when a thermal fraction of unity is assumed; (4) cosmic SFRD when the thermal fraction predicted from synchrotron SFRs is used (Section \ref{sec:discussion_sfrs}).}

\end{deluxetable*}


\section{Conclusions}
\label{sec:conclusion}

We have performed a multi-frequency radio stacking analysis using deep VLA observations across the well-studied COSMOS and GOODS-N fields in order to investigate the shape of the radio spectrum of faint star-forming galaxies. The deep 34\,GHz observations from the COLD$z$ survey form the foundation of this work, and are augmented by deep archival data at 1.4, 3, 5, and 10\,GHz. We construct three mass-complete bins from near-infrared selected galaxy catalogs across COSMOS and GOODS-N, and remove sources that are unlikely to be star-forming based on their optical/near-IR colors and radio emission. We stack at the known positions of the star-forming galaxies at all available frequencies, and decompose the resulting radio spectra into their free-free and synchrotron components.

We detect stacked 34\,GHz emission in the medium ($>10^{9.5}\,\rm{M}_\odot$) and high-mass ($>10^{10}\,\rm{M}_\odot$) bins, and place upper limits on the radio luminosity in the low-mass bin ($>10^{9}\,\rm{M}_\odot$; Figures \ref{fig:stacked_cosmos} and \ref{fig:stacked_goodsn}). Surprisingly, the fitted fractional contribution of free-free emission to the total radio emission at 34\,GHz -- the thermal fraction -- is a factor of $\sim1.5-2$ lower compared to the canonically assumed model for the radio SED (M82-like; Figure \ref{fig:stacked_params}). However, in all cases the stacked 34\,GHz luminosities are consistent with the predicted radio luminosity from free-free emission, when assuming the star formation rates derived from optical-infrared data for the galaxies in the parent catalogs. This points towards a deficit in synchrotron emission at high frequencies (rest-frame $60 - 90\,$GHz), while the contribution from free-free emission is as expected. Accordingly, this implies a high thermal fraction of $f_\text{th} \sim 0.8$ in this frequency range.

Such a synchrotron deficit can plausibly be the result of synchrotron aging of high-energy cosmic rays. Upon combining the radio continuum data across COSMOS and GOODS-N, we fit a more complex model to the radio spectrum including a spectral break. While a precise break frequency can not be robustly ascertained, a break at rest-frame $\nu\gtrsim15\,$GHz is favored. We supplement this analysis with realistic simulations of mock radio spectra, and verify that a spectral break at a rest-frame frequency of $\nu_b \sim 15 - 25\,$GHz, in combination with a typical thermal fraction of $f_\text{th}(1.4\,\text{GHz}) = 0.10$, can explain the observed high-frequency deficit (Figure \ref{fig:mockfitting}). 

Finally, we perform a mean stacking analysis at 34\,GHz, which allows us to constrain the cosmic star-formation rate density with free-free emission at $0.5 \leq z \leq 3.0$ We find good agreement between the constraints from high-frequency radio emission with canonical star-formation rate tracers, including radio synchrotron emission (Figure \ref{fig:SFRD}). This, in turn, demonstrates that free-free emission can reliably be used as a tracer of star-formation in the early Universe. 

Our current analysis remains limited by the 34\,GHz observations, which cover a relatively small area on the sky. In addition, the individual stacks are of modest signal-to-noise, which complicates the spectral decomposition, and hence the determination of thermal fractions and subsequent free-free star-formation rates. Finally, the radio spectra of normal star-forming galaxies may be more complicated than is typically assumed, such that an improved sampling of the radio SED through matched depth multi-frequency observations is required to make further progress. Future radio telescopes, such as the Square Kilometre Array Phase-1 (SKA-1), and in particular the next-generation VLA (ngVLA), will allow for more robust measurements of high-frequency radio emission in distant star-forming galaxies, and will provide more stringent constraints on cosmic star-formation through free-free emission (e.g., \citealt{murphy2015,barger2018}).


\section*{Acknowledgements}
The authors would like to thank the referee for their useful suggestions that improved this paper. The authors would also like to thank F.\ Owen for sharing the 1.4\,GHz radio image of GOODS-N. The National Radio Astronomy Observatory is a facility of the National Science Foundation operated under cooperative agreement by Associated Universities, Inc. H.S.B.A. and J.A.H. acknowledge support of the VIDI research programme with project number 639.042.611, which is (partly) financed by the Netherlands Organization for Scientific Research (NWO). D.R. acknowledges support from the National Science Foundation under grant Nos. AST-1614213 and AST- 1910107. D.R. also acknowledges support from the Alexander von Humboldt Foundation through a Humboldt Research Fellowship for Experienced Researchers. I.R.S. acknowledges financial support from STFC (ST/T000244/1). M.A. acknowledges support from FONDECYT grant 1211951, ``CONICYT + PCI + INSTITUTO MAX PLANCK DE ASTRONOMIA MPG190030'' and ``CONICYT+PCI+REDES 190194''. B.M. acknowledges support from the Collaborative Research Centre 956, sub-project A1, funded by the Deutsche Forschungsgemeinschaft (DFG) -- project ID 184018867. M.T.S. acknowledges support from a Scientific Exchanges visitor fellowship (IZSEZO\_202357) from the Swiss National Science Foundation."

\appendix

\section{Stacking}
\label{app:stacking}

Stacking is the art of co-adding small cutouts around a priori known galaxy positions in order to obtain a census of their typical emission at a different wavelength. Median stacking is often favored over mean stacking in the literature, as it reduces any biases from outliers, and is more representative of the (radio-)undetected population (e.g., \citealt{condon2013} show the mean tends to be skewed towards sources close to the detection threshold). As a result, the median stacked spectrum is not dominated by a few bright, strongly star-forming galaxies, which may show relatively complex radio spectra \citep{tisanic2019,thomson2019}. In addition, a median stacking analysis renders one less susceptible to contamination from radio AGN , which constitute a minority of the population at faint radio fluxes \citep{smolcic2017b,algera2020b}. A further advantage is that, when adopting the median and stacking in luminosity as opposed to flux density, the resulting stacks typically have higher signal-to-noise ratios compared to when the mean is adopted. This is due to the luminosity distance, by which the stacks are multiplied (Equation \ref{eq:lumstacking}), being a strongly increasing function of redshift. In turn, the RMS noise of stacks in units of luminosity density at high redshift is larger than that of the low-redshift population. A simple mean stacking analysis up-weights these noisy stacks, while a noise-weighted mean approach instead down-weights the high-redshift population. Finally, a median stacking analysis allows one to treat individually detected sources and non-detections homogeneously by stacking them together, whereas a mean stacking technique requires additional care to be taken when dealing with bright outliers or neighboring galaxies.

However, one caveat that applies when using the median is typically not addressed: \citet{white2007} show that, given a set of point-source flux densities $S_i$ in an image with local RMS-noise $\sigma_i$, the stacked median $\tilde{S}_\text{stack}$ does not necessarily represent the true sample median $\tilde{S}_\text{true}$. In particular, when $S_i \ll \sigma_i$, the stacked median tends towards the true sample mean, i.e., $\tilde{S}_\text{stack} \rightarrow \overline{S}_\text{true}$. In the opposite scenario, where $S_i \gg \sigma_i$, the stacked median tends towards the true sample median $\tilde{S}_\text{stack} \rightarrow \tilde{S}_\text{true}$. In practice, the typical flux density of the sources is likely to be similar to the RMS, i.e., $S_i \sim \sigma_i$: if the sources are much brighter, stacking is likely unnecessary, while if the sources are substantially fainter, a significant number of galaxies is required to get a stacked detection.\footnote{For example, if the median flux density is just $0.1\sigma_i$, this requires $\sim10^{3}$ sources to be stacked for a detection at $\text{S/N}\approx 3$, assuming that the noise decreases as $1/\sqrt{N}$. In the stacking analysis in this work, a sample of a $\sim60 - 600$ galaxies is typical (Table \ref{tab:binning}).} This, in turn, implies that the stacked median will probe a value somewhere between the true median and true mean, complicating its interpretation. For a typical distribution of flux densities the mean will exceed the median. As a result, when performing a stacking analysis, the median is ``boosted'' compared to the true sample median. The mean does not suffer from this difficulty, and needs not be deboosted, but is instead subject to the previously mentioned drawbacks.

What truly complicates this picture, however, is that the extent to which the median is boosted depends on the typical ratio $S_i / \sigma_i$. While the (local) noise properties of the image, $\sigma_i$, are known, the typical flux densities $S_i$ of course are not -- otherwise, one would not be stacking! Nevertheless, correcting for median boosting is particularly important in the multi-frequency stacking analysis we perform in this work: radio sources typically are fainter at higher rest-frame frequencies, and the radio maps used for stacking are of varying depths. As a result, the level of boosting is dependent on the properties of the individual radio images, and hence stacked flux densities cannot directly be compared, as they do not trace the quantity we would like to compare, $\tilde{S}_\text{true}$. What is needed, then, is an estimate of the boosting factor $f_\text{boost} = \tilde{S}_\text{stack} / \tilde{S}_\text{true}$, which can be used to correct stacked flux densities back to the true sample median.

In order to estimate the boosting factor, we perform a stacking analysis on simulated sources, for which the true sample median and mean are known a priori. Since the level of median boosting depends on the quantity $S_i / \sigma_i$, it is crucial that the distribution of flux densities $S_i$ be realistic. For that reason, we generate mock sources by assigning radio flux densities to galaxies in the COSMOS2015 or 3D-HST catalogs. For each set of mass-complete bins in the COSMOS and the GOODS-N field, we randomly draw the redshift $z$ and stellar mass $M_\star$ for a large number of galaxies within the corresponding catalogs. We then adopt the main-sequence from \citet{schreiber2015} to assign these mock galaxies a star-formation rate, and include a realistic scatter of $0.25\,$dex. We subsequently convert these star formation rates to radio luminosities $L_{1.4}$ at rest-frame 1.4\,GHz via the far-infrared/radio correlation from \citet{bell2003}, including their measured scatter of $0.26\,$dex about the correlation. We then randomly draw radio spectral indices from a normal distribution with a mean of $-0.70$ and a scatter of $0.30$, and use these to calculate the flux densities of each of the mock sources. We assign these to unresolved mock sources which are subsequently inserted into the residual radio maps at randomly selected positions, using a Gaussian beam matching the resolution and position angle of the restoring clean beam. We opt for inserting mock sources into the residual images as to ensure we are not artificially approaching the confusion limit, but are still including realistic noise properties in our analysis. We further include small offsets in RA and Dec, normally distributed around zero, with a scatter of $0\farcs30$, between the true and catalogued mock sources positions, in order to capture realistic spatial offsets between sources. Finally, we stack the mock sources, and compare their input and output median luminosity densities. The ratio of these two quantities then defines the boosting factor, which is used to correct the luminosities probed in the real stacks. 

\begin{figure}[!t]
    \centering
    \includegraphics[width=0.5\textwidth]{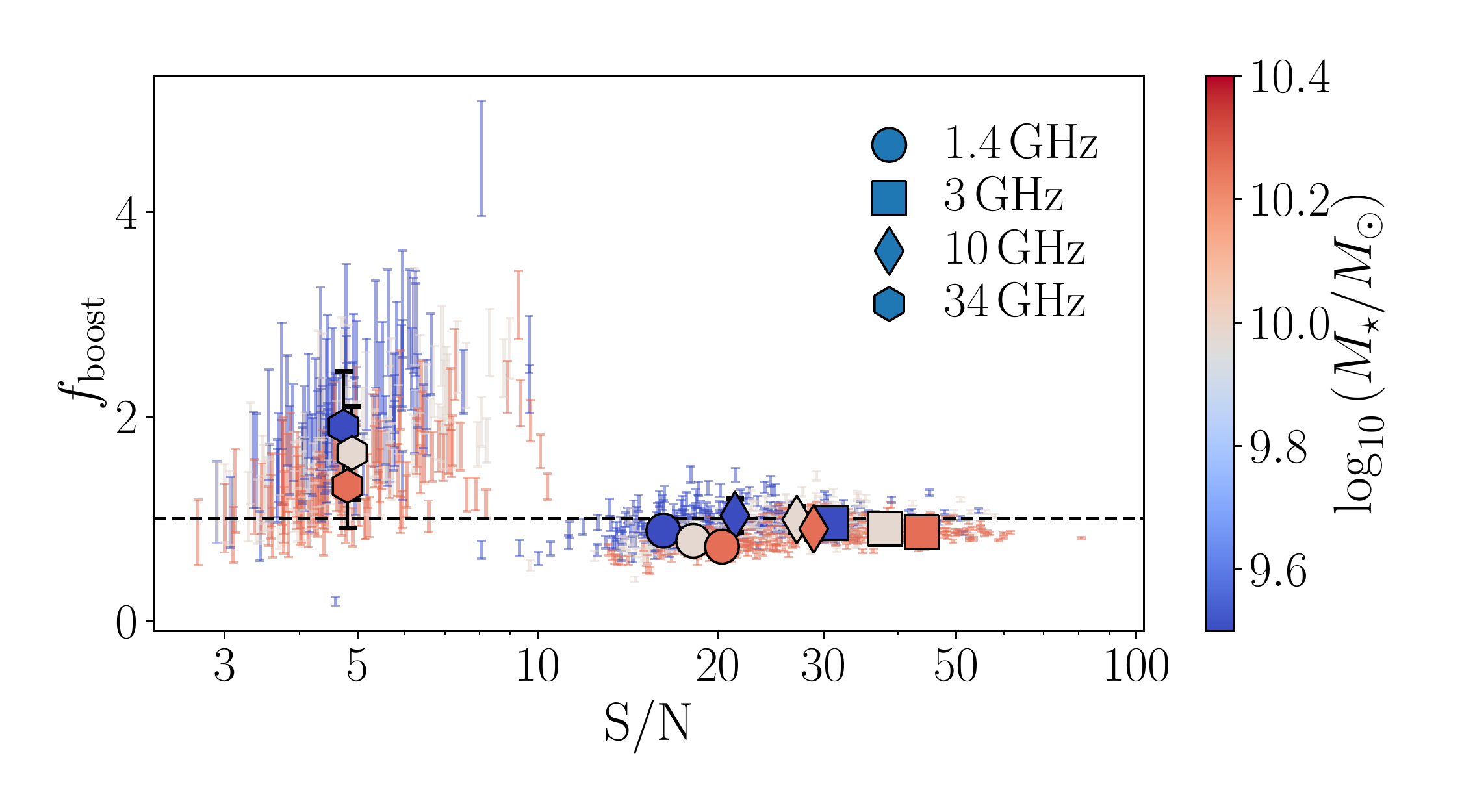}
    \caption{Median boosting as a function of S/N for stacks of simulated sources drawn from the bins defined in Table \ref{tab:binning}. Points are colored by their stellar mass -- at fixed redshift, lower mass galaxies are typically fainter, and hence will show a greater level of median boosting. The correction factor is around unity at 1.4, 3 and 10\,GHz, but reaches $f_\text{boost} = 1.9$ ($\sigma = 0.5$) for the low-mass bin at 34\,GHz. All median-stacked flux densities presented in this work are corrected for boosting.}
    \label{fig:fluxboosting}
\end{figure}

As an example, we show the level of boosting for the COSMOS2015 catalog (Section \ref{sec:results_optical}) in Figure \ref{fig:fluxboosting}. The correction factor is typically around unity at high-S/N, and may in fact be slightly lower than one when the peak flux underestimates the true flux density of the stacked mock sources. However, for low-mass galaxies at 34\,GHz, which on average have lower star-formation than their high-mass counterparts and hence a lower typical $S_{i} / \sigma_{i}$, the boosting correction may reach up to $f_\text{boost} \approx 2$. All median-stacked flux densities and spectral luminosities presented in this work are corrected for boosting, based on realistic simulations such as the COSMOS one presented here. The boosting factors for the stacks in Figures \ref{fig:stacked_cosmos} and \ref{fig:stacked_goodsn} are additionally reported in Table \ref{tab:stacking}. The uncertainty on the stacked luminosities further includes the spread across the recovered boosting corrections.

\section{The COSMOS $z=5.3$ Protocluster}
\label{app:protocluster}

The COLD$z$ COSMOS observations were designed to overlap with a prominent $z=5.3$ protocluster, of which AzTEC-3 is the brightest member \citep{capak2011,riechers2014}. While this source is individually detected in the COLD$z$ 34\,GHz observations, an additional 9 protocluster members remain undetected at 34\,GHz yet are observed in Lyman-$\alpha$ emission in VLT/MUSE spectroscopy (Guaita et al., in preparation). We median stack on all members, excluding AzTEC-3, at 1.4, 3, 10 and 34\,GHz -- in flux density as opposed to luminosity, as all sources lie at the same approximate redshift. However, we find no detection in the stacks at any frequency, and can therefore only place upper limits on the typical high-frequency continuum emission from the protocluster galaxies. At 10 and 34\,GHz, respectively, we place $3\sigma$ upper limits of $S_{10} \lesssim 0.58\,\mu\text{Jy\,beam}^{-1}$ and $S_{34}\lesssim1.7\,\mu\text{Jy\,beam}^{-1}$.

The 10\,GHz stack probes a rest-frame frequency of $\nu'\approx63\,$GHz at $z=5.3$. As such, it directly places constraints on the typical level of free-free emission in the protocluster members. The upper limit for the flux density at 10\,GHz translates into a limit on the luminosity density of $L_{\nu'} < 2.8 \times10^{22}\,\text{W\,Hz}^{-1}$. Adopting a thermal fraction of unity and a fixed median boosting factor of $f_\mathrm{boost} = 2$, we determine an upper limit of $\text{SFR} < 90\,M_\odot\,\text{yr}^{-1}$ for the protocluster members. This is consistent with the findings from \citet{capak2015}, who determine typical star-formation rates of $\sim40\,M_\odot\,\text{yr}^{-1}$ for a subset of the protocluster members used for stacking in this work, based on their combined UV and far-infrared emission. While the 34\,GHz data are a factor of $\sim3-4\times$ less sensitive than the 10\,GHz stack, we may expect thermal emission from dust to become important at these frequencies ($\nu' \approx 210\,$GHz). Indeed, \citet{algera2020c} find that the 34\,GHz continuum emission observed for AzTEC-3 is likely dominated by a combination of dust emission and the CO(2-1) emission line. To estimate the contribution from dust emission at 34\,GHz for the protocluster members, we assume a star-formation rate of $\text{SFR} = 90\,M_\odot\,\text{yr}^{-1}$ and a grey body with $\beta = 1.8$ and $T_\text{dust} = 35\,$K for the dust SED (e.g., \citealt{casey2014}). Given that the assumed star formation rate is an upper limit, this translates into an upper limit on the 34\,GHz flux density due to dust of $S_{34}^\text{dust} < 0.9\,\mu\text{Jy\,beam}^{-1}$. In comparison, the combination of free-free and synchrotron emission, assuming a simple M82-like radio spectrum, is expected to contribute only $S_{34}^\mathrm{radio} < 0.3\,\mu\mathrm{Jy\,beam}^{-1}$ at this frequency.


\bibliographystyle{aasjournal}
\bibliography{main}

\end{document}